\documentclass[a4paper,11pt]{article}  
\usepackage{mathptmx}
\usepackage{bm}
\usepackage{berasans}
\usepackage{beramono}

\usepackage[T1]{fontenc}
\usepackage[latin9]{inputenc}
\usepackage{geometry}
\usepackage{color}
\usepackage[english]{babel}
\usepackage{booktabs}
\usepackage{float}
\usepackage{enumerate}
\usepackage{amsthm}
\usepackage{amsmath}
\usepackage{amssymb}
\usepackage{graphicx}
\usepackage{esint}

\usepackage{amsmath,amsfonts}
\usepackage{graphicx}
\usepackage[colorlinks=true, allcolors=blue]{hyperref}
\usepackage[round]{natbib}

\theoremstyle{definition}
\newtheorem{defn}{\protect\definitionname}[section]
\theoremstyle{plain}
\newtheorem{prop}{\protect\propositionname}[section]
\theoremstyle{definition}
\newtheorem{example}{\protect\examplename}[section]
\ifx\proof\undefined\
\newenvironment{proof}[1][\proofname]{\par
	\normalfont\topsep6\p@\@plus6\p@\relax
	\trivlist
	\itemindent\parindent
	\item[\hskip\labelsep
	\scshape
	#1]\ignorespaces
}{%
	\endtrivlist\@endpefalse
}
\providecommand{\proofname}{Proof}
\fi
\theoremstyle{plain}
\newtheorem{lem}{\protect\lemmaname}[section]
\theoremstyle{plain}
\newtheorem{rem}{\protect\remarkname}[section]
\theoremstyle{plain}
\newtheorem{thm}{\protect\theoremname}[section]
\theoremstyle{plain}

\usepackage{indentfirst}

\makeatother

\providecommand{\corollaryname}{Corollary}
\providecommand{\definitionname}{Definition}
\providecommand{\examplename}{Example}
\providecommand{\lemmaname}{Lemma}
\providecommand{\propositionname}{Proposition}
\providecommand{\remarkname}{Remark}
\providecommand{\theoremname}{Theorem}

\title{Asymptotic fractional-order stochastic dominance with bounded relative risk aversion}

\author{\textit{Jiehua Xie$^{\rm a}$,\ \ Liulei Sun$^{\rm b*}$,
		\ \ Wei Zou$^{\rm c}$\thanks{Corresponding			authors.			 \textit{Email-addresses:}l229sun@uwaterloo.ca(Liulei Sun), zouwei@juwp.edu.cn(Wei Zou).}}	\\	\small  \textit{$^{\rm a}$School of Business Administration, Jiangxi University of Water Resources and Electric Power, Jiangxi 330099, P.R.China}\\	\small \textit{$^{\rm b}$Department of Statistics and Actuarial Science, University of Waterloo, Canada}
	\\ \small	  \textit{$^{\rm c}$School of Science, Jiangxi University of Water Resources and Electric Power, Jiangxi 330099, P.R.China}	}

\date{July 2026}

\begin{document}
\maketitle
\begin{abstract}
In this paper, we propose a novel asymptotic fractional-order  stochastic dominance rule for ranking prospects over a sufficiently long investment horizon. The new rule formulates the consensus of decision makers whose
relative risk aversion has a negative lower bound. Under the assumption that returns are lognormally distributed, we establish  equivalent conditions for the proposed rule without imposing the non-negativity constraint on the mean of log-return, a restriction usually required by the existing asymptotic stochastic dominance rules. Furthermore, to enhance the tractability of this asymptotic fractional-order stochastic dominance,  
we propose a variant of asymptotic fractional-order stochastic dominance with bounded relative risk aversion, referred to  as general asymptotic fractional-order stochastic dominance, under an additional condition on decision makers' marginal utilities. We  derive its corresponding equivalent distributional characterizations. The (general) asymptotic fractional-order  stochastic dominance with bounded relative risk aversion  overcomes the shortcomings of the existing asymptotic fractional-order   criterion that the fractional-order parameter has no influence on the equivalent distributional  conditions. Empirical examples further show the advantages of the newly proposed rules for asset selection in long-term investment decisions.

\end{abstract}
%

\section{Introduction}

The increase in life expectancy presents significant challenges not only in finance and  economics, 
but also in the field of  management \citep{2022Levy,2024Understanding,2023Goodman,2025Registered,2025Household}. Within the context of life-cycle economic planning in an aging society, the determination of an optimal long-term portfolio   has gradually emerged as a critical issue \citep{2014Poterba,2024Chambers,2023Levy,2024Levy}. As essential tools in comparing and ranking prospects in decision-making under uncertainty, stochastic dominance rules have
also been employed for optimal portfolio selection in long-term investments. To address the optimal portfolio selection over a sufficiently long investment horizon, \citet{Levy2016Aging} first introduced the concept of asymptotic first-order stochastic dominance (AFSD, for short) by incorporating the effect of the investment horizon into the framework of first-order stochastic dominance (FSD). Using the AFSD rule, \citet{Levy2016Aging} established the optimality of the maximum geometric mean portfolio for the long-term investments. Subsequently, \cite{2025Levy} further examined the maximum geometric mean rule under the constant relative risk aversion utility function.

The AFSD provides a basis for  asset selection that is  applicable to all non-satiable decision makers (DMs) as the investment horizon  approaches infinity. This framework elucidates  how  preferences shape risky choices of DMs in long-term investments. To derive the equivalent conditions for AFSD,  \citet{Levy2016Aging} assumed that returns in each period are independent and identically lognormally distributed. This is a  widely accepted assumption in  financial modeling and risk management   \citep{2017Portfolio,2018Non,Fama2018Long,2023Asymptotic,2021Stocks}. Based on this assumption, \citet{Huang2020Comment} established the necessary and sufficient moment conditions for AFSD. Furthermore,  \citet{Levy2020Aging} provided empirical evidence supporting the assertion that, for non-satiable DMs, a portfolio with a higher geometric mean return stochastically dominates all others as the investment horizon  tends to infinity. Although AFSD offers a valuable framework for addressing investment decisions in aging societies, its associated distributional conditions are often excessively restrictive for practical application. To address this limitation, \citet{Huang2020Operational} extended the concept to asymptotic second-order stochastic dominance (ASSD), which serves as a ranking rule  for all non-satiable and risk-averse DMs when the investment horizon is sufficiently long.  
Under the assumption that returns follow the lognormal distribution, they also obtained corresponding necessary and sufficient distributional conditions for ASSD.  

Within the expected utility framework, AFSD requires the DM's utility function to be non-decreasing, whereas  ASSD imposes the additional condition of concavity, thereby requiring the utility function to be both non-decreasing and concave. The distinction between ``non-decreasing'' and ``non-decreasing concave'' is significant, as the latter excludes all utility functions that exhibit local convexities. 
Consequently, the jump from AFSD to ASSD  is notably substantial, since any utility function with convex segments is precluded if a DM aims to apply an asymptotic stochastic dominance rule of degree greater than one in long-term investments. Nevertheless, utility functions with local convexities are capable of capturing  various individual behaviors, such as the simultaneous purchase of insurance and lottery tickets   \citep{Jullien2000Estimating,2018Schneider,xie2022patchwork}, and they have consequently attracted considerable attention in the literature \citep{1989Friedman,1979Kahneman}. To accommodate  utility functions that exhibit  such local convexities, \cite{2026Xie} proposed the asymptotic fractional-order stochastic dominance rule, named as  $1+c$-ASD ($0\le c\le 1$), and established the necessary and sufficient distributional conditions for this fractional-order rule under lognormal assumption. Under the $1+c$-ASD, the portfolio that yields a higher geometric mean return coupled with lower volatility constitutes the dominant portfolios for all non-satiable DMs whose utility functions exhibit local convexities. 

However, the fractional-order parameter $c$ does not influence the distributional conditions of  the $1+c$-ASD rule, implying that for all $0<c<1$, the corresponding distributional conditions remain identical. This invariance appears unrealistic and may  potentially weaken the main contribution of the $1+c$-ASD, which aims to establish a consensus distribution ranking by excluding some non-satiable DMs. For example, consider two assets, $A$ and $B$, in long-term investment decisions, and suppose that the daily returns of assets follow lognormal distributions. The parameters of the corresponding daily log-returns for each asset are presented in Table  \ref{tab1+}. According to the equivalent conditions established in \citet{Levy2016Aging} and \cite{2026Xie},  assets $A$ and $B$ can not be ranked over a sufficiently long investment horizon by AFSD. However, asset $B$ asymptotically  dominates asset $A$ by $1+c$-ASD for all $0<c\le1$. Since the fractional-order parameter $c$ does not influence the equivalent conditions of the $1+c$-ASD rule, a substantial discontinuity arises in the asymptotic dominance relation when  moving from $c=0$ to $c>0$. Such a discontinuity may be unrealistic. 

	\begin{table}[h]
	\centering
	\caption{The parameters of daily log-returns for two assets.}
	\label{tab1+}
	\begin{tabular}{lccc}
		\toprule
		Asset & Mean  $\mu$ &  Variance $\sigma^2$ & $\mu+\sigma^2/2$ \\
		\midrule
	$A $       & 3 & 0.11 & 3.055 \\
	$B$      & 3.01 & 0.1 & 3.06  \\
		\bottomrule
	\end{tabular}
\end{table}

Motivated by the above insights, this paper proposes a novel asymptotic fractional-order stochastic dominance rule for ranking assets in long-term investments, addressing the key  shortcoming of the existing rules that the fractional-order parameter does not influence the distributional conditions. Firstly,  within the expected utility framework, we assume that the DM's Arrow-Pratt index of relative
risk aversion is no less than $-(1/\ell-1)$, where $0 < \ell \le 1$. This assumption establishes a continuum that interpolates between first-order and  second-order stochastic dominance. We refer to the resulting distribution 
ranking rule for such DMs as $(1+\ell)$th-order
stochastic dominance, denoted as  $1+\ell$-SD. We examine the  equivalent distributional conditions for $1+\ell$-SD. Subsequently, we incorporate the influence of the investment horizon into the framework of  $1+\ell$-SD.  Under the assumption of lognormally distributed returns, we derive   necessary and sufficient  distributional conditions for  $1+\ell$-SD  when the investment horizon is finite. Extending this framework to infinite horizons,  we introduce the  asymptotic stochastic dominance of degree $1+\ell$ ($1+\ell$-ASD) as a rule for ranking assets in  long-term investments. We establish the necessary and sufficient distributional conditions for $1+\ell$-ASD, which depend on the fractional-order parameter $\ell$. Furthermore, to enhance the tractability of $1+\ell$-ASD, we place an additional condition on DMs' marginal utilities and introduce the general  asymptotic stochastic dominance of degree $1+\ell$ (general  $1+\ell$-ASD). We also provide the equivalent distributional conditions for general $1+\ell$-ASD, which relax the volatility constraint on the $1+\ell$-ASD. Finally, we investigate the practical applicability of the newly proposed $1+\ell$-ASD and general $1+\ell$-ASD rules. In the context of long-term investment decisions, the usefulness of these proposed rules for asset selection is illustrated through empirical  analysis. The empirical results show that the (general) $1+\ell$-ASD rule offers a tractable framework for   pairwise comparisons of individual assets, thereby providing reasonable decision for long-term investment choices.

 This study contributes to introducing an alternative fractional-order stochastic dominance rule and providing its equivalent distributional conditions. \cite{Mullerd2017Between} and \cite{Huang2020Fractional} highlighted a significant limitation of  existing stochastic dominance rules: the integer degree scale is too coarse. To address this, they developed fractional-order stochastic dominance rules that cover  preferences from first-order to second-order stochastic dominance (SSD), by imposing constraints on the ratio of marginal utilities and on the degree of absolute risk aversion, respectively. In contrast to these approaches of \cite{Mullerd2017Between} and \cite{Huang2020Fractional}, our dominance criterion, that is, the $1+\ell$-SD rule, captures the consensus of DMs whose relative risk aversion has a negative lower bound. 
 
 Furthermore, our study contributes to the development of a novel asymptotic fractional-order stochastic dominance rule between AFSD and ASSD, which is achieved by incorporating the effect of the investment horizon into the framework of $1+\ell$-SD. To bridge the gap between ``non-satiable'' and ``non-satiable and risk-averse'' DMs, \cite{2026Xie} first introduced a continuum of asymptotic stochastic dominance rules, referred to as $1+c$-ASD, which encompasses both the AFSD rule introduced by \citet{Levy2016Aging} and the ASSD rule presented by  \citet{Huang2020Operational}. The  $1+c$-ASD is capable of capturing the preferences of DMs whose utility functions exhibit local convexities, such as these characterized by $S$-shaped reference-dependent utility functions \citep{1989Friedman,1979Kahneman}. The newly proposed  $1+\ell$-ASD is likewise applicable to DMs who are predominantly risk-averse but do not categorically reject all forms of risk. In contrast to \cite{2026Xie}, who placed  constraints
 on the ratio of marginal utilities, the present study imposes  a lower bound for the relative risk aversion. It is worth noting that the proposed $1+\ell$-ASD overcomes the key limitation of $1+c$-ASD that the fractional-order parameter $c$ does not influence the equivalent distributional conditions. The distributional conditions of $1+\ell$-ASD form a continuum ranging from those of AFSD to those of ASSD, thereby addressing the marked discontinuity in the asymptotic dominance relation under $1+c$-SD rule when moving from $c=0$ to $c>0$. Consequently, the $1+\ell$-ASD can more accurately rank assets over a sufficiently long investment horizon. 
 
In addition,  under the assumption that returns follow lognormal distributions, we establish equivalent conditions for the newly proposed $1+\ell$-ASD and general $1+\ell$-ASD rules without imposing the non-negativity constraint on the mean of log-return, a restriction usually required by the existing asymptotic stochastic dominance rules, such as \cite{Huang2020Operational} and \cite{2026Xie}. Consequently, our conditions are applicable to a broader range of scenarios.
 
 The remainder of this paper is organized as follows. Section 2 introduces the definition of $1+\ell$-SD and establishes its equivalent distributional conditions. We also incorporate the influence of the investment horizon into the framework of  $1+\ell$-SD in Section 2. Under the assumption of lognormally distributed  returns and for a finite investment horizon,  the necessary and sufficient conditions for $1+\ell$-SD are  presented. Section 3  defines $1+\ell$-ASD rule and derives its equivalent distributional conditions. In Section 4,  we introduce the general $1+\ell$-ASD  rule and provide its corresponding equivalent distributional conditions. Section 5 presents an empirical analysis to show the practical applicability of $1+\ell$-ASD and general $1+\ell$-ASD rules in long-term  investments. Concluding remarks are presented in Section 6. Some preliminary results and all proofs are put in Appendix.

\section{Fractional-order stochastic dominance with bounded relative risk aversion}

\subsection{ Stochastic dominance of degree 1+$\ell$}

Let $U$ denote the utility function of a DM, which is assumed to be  
continuous and piecewise differentiable. Denote a lottery by $X$ and its expected utility by $\mathbb{E}[U(X)]$. Throughout this paper, we consider  lotteries with the support $(0,+\infty)$.

To accommodate utility functions that exhibit local convexities, we consider the following class of utility functions. 


\begin{defn}\label{gammaU}
	For $0< \ell\le 1$, let $\mathcal{U}_{1+\ell}$ be the class of utility functions $U$ such that 
	\begin{equation}\label{eq:gammaU}
	U'(x)\ge 0 \ \ {\rm and}\ \ U'(x)/x^{1/\ell-1}\ \ \text{is non-increasing on}\ \ (0,\infty). 
	\end{equation}
\end{defn}

~

 Denote the Arrow-Pratt index of relative risk aversion associated with $U$ by 
$$R_{U,2}(x)=-\frac{xU''(x)}{U'(x)}.$$
It is worth noting that when $U$ is twice continuously differentiable and strictly increasing, i.e.,  $U'>0$, for each $0< \ell\le 1$, $U\in \mathcal{U}_{1+\ell}$ is equivalent to that $U'(x)>0$ and 
$$R_{U,2}(x)\ge -\big(\frac{1}{\ell}-1\big)\ \ \text{for all}\ \ x> 0.$$
Therefore, the condition \eqref{eq:gammaU} on marginal utilities is equivalent to imposing a lower bound on the index of relative risk aversion $R_{U,2}(x)$, where the bound is determined by the parameter $\ell$. 

As $\ell$ approaches $0$, the constraint on the index of relative risk aversion diminishes, and in this  limiting case, the class $\mathcal{U}_{1}$ encompasses all non-satiable DMs. Conversely, as $\ell$ tends to $1$, the degree of risk-seeking behavior becomes increasingly restricted, and in this  limiting case, only non-satiable and risk-averse DMs are included in $\mathcal{U}_{2}$. For values of $\ell$ within the interval (0,1),  preferences may admit a limited degree of risk-seeking behavior, and the class $\mathcal{U}_{1+\ell}$ contains a variety of utility functions that exhibit local convexities.

 
\begin{example}
 The convex power utility $U_1(x)=x^{\nu}$ ($\nu>$ 1)  and the convex exponential utility $U_2(x)=e^{\lambda x}$ 
 ($\lambda>$ 0)
are commonly employed in study of risk-seeking behavior
\citep{1987A,2013Even,2008Risk,2015Risk}. It can be
shown that  $U_1(x)$, defined on $(0, \infty)$, belongs to $\mathcal{U}_{1+\ell}$ for all $0<\ell\le 1/\nu$, and $U_2(x)$, defined on the interval $(0,1]$, belongs to the class 
$U\in \mathcal{U}_{1+\ell}$ for all $0<\ell\le 1/(1+\lambda)$.
\end{example}

For two lotteries, $X$ and $Y$, with cumulative distribution functions (CDFs) $F$ and $G$, respectively, using the class of utility functions  $\mathcal{U}_{1+\ell}$, we define the \textit{stochastic dominance relation of degree $1+\ell$} ($1+\ell$-SD) between $X$ and $Y$.

\begin{defn}\label{cFSD}
	For $0< \ell\le 1$, $X$ dominates $Y$ by  1+$\ell$-SD, denoted $X\succsim_{1+\ell-\textit{SD}}Y$ if and only if 
	\begin{equation}\label{eq:cFSD}
	\mathbb{E}[U(X)] \ge \mathbb{E}[U(Y)]\ \ \text{for all}\ \ U\in \mathcal{U}_{1+\ell},
	\end{equation}
	and there exists some $U\in\mathcal{U}_{1+\ell}$ for which the inequality is strict. 
\end{defn}

Since for each $0< \ell\le 1$, $U\in \mathcal{U}_{1+\ell}$ corresponds to imposing a lower bound $-(1/\ell-1)$ on the index of relative risk aversion $R_{U,2}(x)$, then $1+\ell$-SD rule is also referred to as  \textit{stochastic dominance with bounded relative risk aversion}.

\begin{rem}\label{rem:equ+} 
Notice easily that $\mathcal{U}_{1+\ell_2}\subset \mathcal{U}_{1+\ell_1}$ when $0\le \ell_1<\ell_2\le 1$. Consequently, if $0\le \ell_1<\ell_2\le 1$ 	and there exists some $U\in\mathcal{U}_{1+\ell_2}$ for which the inequality \eqref{eq:cFSD} is strict, then
 $$X\succsim_{1+\ell_1-\textit{SD}}Y\Rightarrow X\succsim_{1+\ell_2-\textit{SD}}Y.$$
\end{rem}

\cite{Mullerd2017Between} developed a continuum of stochastic dominance rules by bounding the ratio of marginal utilities, that is, $0\le \gamma U'(y)\le U'(x)$ for all $x\le y$, where $0\le \gamma\le 1$. \cite{Huang2020Fractional} proposed a fractional-order stochastic dominance rule by imposing a lower bound on the absolute risk aversion, i.e., $U'(x)>0$ and  $\frac{U''(x)}{U'(x)}\ge-\big(\frac{1}{c}-1\big)$,  $0<c\le 1$. In contrast to the approaches of \cite{Mullerd2017Between} and \cite{Huang2020Fractional}, the $1+\ell$-SD rule imposes a lower bound on the relative risk aversion.

The following theorem
provides the equivalent distributional condition for 1+$\ell$-SD.

\begin{thm}\label{Cthm} Denote the CDFs of  $X$ and $Y$ by $F$ and $G$, respectively. 
	For $0< \ell\le 1$, $X$ dominates $Y$ by $1+\ell$-SD  if and only if  
	\begin{equation}\label{cfracSD}
		\int^x_{0}\big(G(t)-F(t)\big){\rm d}h_{\ell}(t)\ge 0\ \ \textit{for all}\ \  x> 0,
	\end{equation}
and there exists some $x_0>0$ for which the inequality is strict, where $h_{\ell}(x)=x^{1/\ell}$, $0<\ell\le 1$. 
\end{thm}

The condition for 1+$\ell$-SD becomes particularly simple when the  distributions are single crossing.

\begin{prop}\label{prop-Cthm}
	Suppose that there exists a single crossing point $x_0\in(0,\infty)$ such that $F(x)\le G(x)$ for $x\le x_0$ and $F(x)\ge G(x)$ for $x\ge x_0$. For $\ell\in(0,1]$,  $X\succsim_{1+\ell-\textit{SD}}Y$ if and only if 
	\begin{equation}\label{eq-Cthm}
	\int_0^{x_0}\big(G(t)-F(t)\big){\rm d} h_{\ell}(t)\ge \int^{\infty}_{x_0}\big(F(t)-G(t)\big){\rm d} h_{\ell}(t) \ \ {\rm and}\ \  \int_0^{x_0}\bigl(G(t)-F(t)\bigr)dh_\ell(t)>0.
	\end{equation}
\end{prop}

\begin{example}
	Consider two lotteries $X$ and $Y$ with CDFs
	\begin{equation*}F(x)
=\left\{\begin{array}{ll}
0,  & \ \text{if}\ x< 0,\\
x,  & \ \text{if}\ 0\le x< 1,\\
1,  & \ \text{if}\ x\ge 1,
\end{array}\right.\ \ {\rm and}\ \ \ 
G(x)
=\left\{\begin{array}{ll}
0,  & \ \text{if}\ x< 0,\\
p,  & \ \text{if}\ 0\le x< 1,\\
1,  & \ \text{if}\ x\ge 1,
\end{array}\right. \ \ {\rm where}\ \ 0<p<1,
\end{equation*}
respectively. Note that,  $F(x)\le G(x)$ for $x\le p$ and $F(x)\ge G(x)$ for $x\ge p$. Simple manipulations yield that for all $\ell\ge 1/p-1$,  $\int_0^{p}\big(G(t)-F(t)\big){\rm d} h_{\ell}(t)\ge \int^{\infty}_{p}\big(F(t)-G(t)\big){\rm d} h_{\ell}(t)$. Thus, it follows from Proposition \ref{prop-Cthm} that for all $\ell\ge 1/p-1$, $X\succsim_{1+\ell-\textit{SD}}Y$. 
\end{example}

\subsection{Stochastic dominance of degree $1+\ell$ over a finite investment horizon}

In this subsection,  we incorporate the influence of the investment horizon into the framework of  $1+\ell$-SD. Assume that the random walk hypothesis holds, i.e., the changes in share prices are serially independent. Under this assumption, a DM's terminal wealth $W_T$ at time $T$ is given by 
\begin{equation}\label{eq:WT}
	W_T=W_0\prod_{k=1}^T(1+R_k),
\end{equation}
where $R_k$, $k\in\{1,\ldots,T\}$ denotes the rate of return in the period $k$ and  $W_0$ represents the DM's initial wealth. Taking logarithms of \eqref{eq:WT}, it yields that 
\begin{equation}\label{eq:logWT}
	\log(W_T)=\log(W_0)+\sum_{k=1}^T\log(1+R_k).
\end{equation}
For simplicity, denote $1+R_k=\tilde{R}_k$ and let  $W_0=1$. Equality  \eqref{eq:logWT} can then be rewritten as 
\begin{equation*}\label{eq:logWT+}
	\log(W_T)=\sum_{i=1}^T\log(\tilde{R}_i).
\end{equation*}
Throughout this paper, we assume that for any $U\in\mathcal{U}_{1+\ell}$ and any finite horizon $T$, $\mathbb{E}[|U(W_T)|]<\infty$.

To better understand the dominant strategy over an investment horizon, \citet{Levy2016Aging} imposed the assumption of  lognormally  distributed portfolio returns, supposing that $\tilde{R}_k$, $k\ge 1$ are independently and identically distributed (i.i.d.) and that, for each $k$, $\log(\tilde{R}_k)$ follows a normal distribution, i.e., $\log(\tilde{R}_k)\sim N(\mu,\sigma^2)$, where $\mu$ and $\sigma^2$ denote the mean and variance, respectively. It then follows that $\log(W_T)$ is normally distributed as $N(T\mu, T\sigma^2)$. Consequently, $W_T$ follows a lognormal distribution $\Lambda\big(\mathbb{E}[W_T], {\rm Var}(W_T)\big)$ with parameters $\mathbb{E}[W_T]=e^{T(\mu+\sigma^2/2)}$ and ${\rm Var}(W_T)=e^{T(2\mu+\sigma^2)}(e^{T\sigma^2}-1)$. The assumption of  lognormally distributed returns has become a  widely accepted condition in financial modeling.    \citet{Levy2020Aging} and \citet{Huang2020Comment,Huang2020Operational} also adopted this assumption in their respective studies, and the present analysis likewise relies on it. Accordingly, the CDFs $F_T$ and $G_T$ of the terminal wealth $W_T$ at time $T$ are two lognormal distribution functions with means  $e^{T(\mu_F+\sigma_F^2/2)}$ and $e^{T(\mu_G+\sigma_G^2/2)}$, respectively. Moreover,  $\mathbb{E}_{F}[\log(W_T)]=T\mu_F$, $\mathbb{E}_{G}[\log(W_T)]=T\mu_G$,  ${\rm Var}_{F}\big(\log(W_T)\big)=T\sigma^2_F$  and ${\rm Var}_{G}\big(\log(W_T)\big)=T\sigma^2_G$, respectively.

In the following theorem, for a finite investment horizon $T$, we provide the necessary and sufficient conditions for the dominance of $F_T$ over $G_T$ within the class $\mathcal{U}_{1+\ell}$ under the assumption of lognormal distributions. 

\begin{thm}\label{fractional} Suppose that $F_T$ and $G_T$ follow lognormal distributions with means  $e^{T(\mu_F+\sigma_F^2/2)}$ and $e^{T(\mu_G+\sigma_G^2/2)}$, respectively. 
For a finite investment horizon $T$ and a constant $0< \ell\le 1$,  $F_T\succsim_{1+\ell-\textit{SD}}G_T$ if and only if  
\begin{equation}\label{condition1}
	\mu_F+\frac{1}{2\ell}\sigma^2_F\ge \mu_G+\frac{1}{2\ell}\sigma^2_G\ \ \  \text{and}\ \ \ \sigma_F\le \sigma_G,
\end{equation}
and at least one of the above inequalities is strict.
\end{thm}

Table \ref{Table:ST} summarizes the equivalent distributional conditions for FSD, SSD, $1+c$-SD \citep{2026Xie} and $1+\ell$-SD under the assumption of lognormally distributed returns with a finite investment horizon $T$. It is observed that the equivalent distributional conditions for FSD, SSD, and $1+\ell$-SD are independent of the investment horizon $T$, whereas those for $1+c$-SD depend explicitly on  $T$.

\begin{table}[h]\footnotesize
	\centering
	\caption{Moment conditions for the dominance relation of $F_T$ over $G_T$  under the assumption of lognormally distributed returns with a finite investment horizon $T$.}
	\label{Table:ST}
	\resizebox{\textwidth}{!}{%
	\begin{tabular}{*{111}{l}}
		\hline\hline\noalign{\smallskip}
		Stochastic dominance & Moment conditions \\ 
		\noalign{\smallskip}\hline\noalign{\smallskip}
	FSD & $\mu_F>\mu_G$ and $\sigma_F=\sigma_G$ & \citet{Levy2016Aging} \\
	SSD & $\mu_F+\sigma^2_F/2\ge\mu_G+\sigma^2_G/2$ and $\sigma_F\le\sigma_G$ with at least one inequality being strict & \citet{Levy2016Aging} \\
		$1+c$-SD ($0\le c\le1$) & $\mu_F\ge\mu_G$ and $\sigma_F\le\sigma_G$ with at least one inequality being strict, & \cite{2026Xie}\\
&  and $e^{(\mu_F+\sigma^2_F/2)T}-e^{(\mu_G+\sigma^2_G/2)T}\ge(1-c)\big\{e^{(\mu_F+\sigma^2_F/2)T}\Phi\big(\sqrt{T}(m+(\sigma_G-\sigma_F)/2)\big)-$ & \\ 
& $e^{(\mu_G+\sigma^2_G/2)T}\Phi\big(\sqrt{T}(m-(\sigma_G-\sigma_F)/2)\big)\big\}$, where  $m=(\mu_F-\mu_G)/(\sigma_G-\sigma_F)-$  &  \\
		&  $(\sigma_F+\sigma_G)/2$ 
for $\sigma_F<\sigma_G$, otherwise $m=+\infty$ &  \\ 
	$1+\ell$-SD ($0<\ell \le1$) & $\mu_F+\frac{1}{2\ell}\sigma^2_F\ge \mu_G+\frac{1}{2\ell}\sigma^2_G$ and $\sigma_F\le \sigma_G$ with at least one inequality being strict & Theorem \ref{fractional}\\
		\noalign{\smallskip}\hline
	\end{tabular}
}
\end{table}

Consider the limiting case $\ell=0$. 
Notice that the condition $\mu_F+\frac{1}{2\ell}\sigma^2_F\ge \mu_G+\frac{1}{2\ell}\sigma^2_G$ in Theorem \ref{fractional} can be rewritten as $\mu_F-\mu_G\ge\frac{1}{2\ell}(\sigma^2_G-\sigma^2_F)$. 
If $\sigma_F<\sigma_G$, the above inequality implies  $\mu_F-\mu_G>+\infty$ as $\ell\rightarrow 0$, which leads to a contradiction. Consequently, it must hold that $\sigma_F=\sigma_G$, and thus $\mu_F>\mu_G$ in the case $\ell=0$. 
Summarizing the above results,  Theorem \ref{fractional} shows that $F_T$ dominates $G_T$ for a finite horizon $T$ within $\mathcal{U}_{1}$ if and only if  $\sigma_F=\sigma_G$ and  $\mu_F>\mu_G$, which coincide with the  moment conditions for FSD presented in \citet{Levy2016Aging}. 

When $\ell=1$, Theorem \ref{fractional} shows that $F_T$ dominates $G_T$ for a finite horizon  $T$ within $\mathcal{U}_{2}$ if and only if     $\sigma_F\le\sigma_G$ and $\mu_F+\sigma^2_F/2\ge \mu_G+\sigma^2_G/2$, where at least one of the inequalities is strict. This result coincides with the conditions for SSD in \citet{Levy2016Aging}.

Therefore, the moment conditions for FSD and SSD represent  two extremes of  the moment condition for  $1+\ell$-SD. Theorem \ref{fractional} consequently establishes the equivalent conditions under which a lognormal  distribution $F_T$ with mean  $e^{(\mu_F+\sigma^2_F/2)T}$ stochastically  dominates another  lognormal distribution $G_T$ with mean  $e^{(\mu_G+\sigma^2_G/2)T}$ for a finite investment horizon $T$,  across a continuum of stochastic dominance rules ranging from FSD to SSD.  The $1+c$-SD rule ($0\le c\le1$) is also a continuum of stochastic dominance rules. However, the moment condition for  $1+c$-SD is more complex to  verify than that for  $1+\ell$-SD. 

\section{Asymptotic stochastic dominance of degree $1+\ell$}

This section considers the asymptotic case as the investment horizon $T\rightarrow\infty$. We develop a continuum of asymptotic stochastic dominance rules and, under the assumption of  lognormally  distributed returns, derive their corresponding equivalent conditions.  

In the remainder of this paper, we impose  the following assumption on the distribution of the terminal wealth $W_T$.

\noindent {\bf Assumption I.}\ \ Let $W_T$ be a lognormal distribution $\Lambda\big(\mathbb{E}[W_T], {\rm Var}(W_T)\big)$ with parameters $\mathbb{E}[W_T]=e^{T(\mu+\sigma^2/2)}$ and ${\rm Var}(W_T)=e^{T(2\mu+\sigma^2)}(e^{T\sigma^2}-1)$. Assume that 
\begin{equation}\label{a}
	\mu+\frac{\sigma^2}{2\ell}>0,\ \ 0<\ell\le 1.
\end{equation}

\begin{rem}
	The condition \eqref{a} is mild. In fact, for any $\mu$,  it is satisfied for $\ell$ sufficiently close to 0. Moreover, the requirement \eqref{a} is less restrictive than the constraint imposed by   \cite{Huang2020Operational} and \cite{2026Xie}, which requires the non-negativity of  $\mu$, as well as the condition in \cite{Huang2020Comment}, which impose the non-negativity on  $\mu+\frac{\sigma^2}{2}$. 
\end{rem}

At time $T$, 
let $\mathbb{E}_{F}[U(W_T)]$ and $\mathbb{E}_{G}[U(W_T)]$ denote the expected utilities of the terminal wealth $W_T$ under the distributions $F_T$ and $G_T$, respectively. Employing the class of utility functions $\mathcal{U}_{1+\ell}$, we formally define the \textit{asymptotic stochastic dominance of degree $1+\ell$} (1+$\ell$-ASD). 

\begin{defn}\label{AFSD}
	For $0< \ell\le 1$, $F_T$ dominates $G_T$ by  1+$\ell$-ASD, denoted $F_T\succsim_{1+\ell-\textit{ASD}}G_T$, if and only if 
	\begin{equation}\label{eq:AFSD}
\liminf_{T\rightarrow\infty}\big(\mathbb{E}_F[U(W_T)]-\mathbb{E}_G[U(W_T)]\big)\ge 0\ \ \text{for all}\ \ U\in \mathcal{U}_{1+\ell},
	\end{equation}
and there exists some $U\in\mathcal{U}_{1+\ell}$ for which the inequality is strict. 
\end{defn}  

In Definition \ref{AFSD}, we extend the limit $\lim\limits_{T\rightarrow\infty}$ employed  in existing asymptotic stochastic dominance rules, such as AFSD \citep{Levy2016Aging}, ASSD \citep{Huang2020Operational} and $1+c$-ASD \citep{2026Xie}, to the limit inferior  $\liminf\limits_{T\rightarrow\infty}$. Consequently,  our definition of 1+$\ell$-ASD does not require the existence of the limit  $\lim\limits_{T\rightarrow\infty}\big(\mathbb{E}_F[U(W_T)]$
	$-\mathbb{E}_G[U(W_T)]\big)$.

Notice that 
$\ell=0$ corresponds to AFSD \citep{Levy2016Aging}, $\ell=1$ corresponds to ASSD \citep{Huang2020Operational}, and $0<\ell<1$ corresponds to asymptotic dominance relationships that fall  between AFSD and ASSD. The set inclusion  $\mathcal{U}_{2}\subset\mathcal{U}_{1+\ell}\subset\mathcal{U}_{1}$ implies that AFSD$\Rightarrow$$1+\ell$-ASD$\Rightarrow$ASSD,  if there exists some $U\in\mathcal{U}_{2}$ for which the inequality \eqref{eq:AFSD} is strict. Furthermore, the set inclusion $\mathcal{U}_{1+\ell_2}\subset\mathcal{U}_{1+\ell_1}$ for  $0\le \ell_1<\ell_2\le 1$ indicates that $1+\ell_1$-ASD$\Rightarrow$$1+\ell_2$-ASD, if there exists some $U\in\mathcal{U}_{1+\ell_2}$ for which the inequality \eqref{eq:AFSD} is strict.

In the following theorem, we establish equivalent distributional conditions for $1+\ell$-ASD, $0<\ell<1$, under the assumption of lognormally distributed returns. 

\begin{thm}\label{thm:AFSD}
 Assume that $F_T$ and $G_T$ are lognormal distributions with means  $e^{T(\mu_F+\sigma_F^2/2)}$ and $e^{T(\mu_G+\sigma_G^2/2)}$, respectively. For $0< \ell\le 1$ and  $T\rightarrow\infty$, $F_T$ dominates $G_T$ by  $1+\ell$-ASD, if and only if the condition \eqref{condition1} holds. 
\end{thm}

Table \ref{Table:ST2} presents a comparative analysis of the equivalent conditions associated with AFSD, ASSD, $1+c$-ASD ($0<c<1$), and $1+\ell$-ASD ($0<\ell<1$). The comparison results show that, relative to the distributional conditions required for AFSD, the conditions for $1+\ell$-ASD ($0<\ell<1$) are less restrictive, while those for  ASSD are the least strict. In contrast to AFSD, $1+\ell$-ASD ($0<\ell<1$) permits $\sigma_F<\sigma_G$ to ensure that $F_T$ is preferred to $G_T$ as $T\rightarrow \infty$ for all non-decreasing utility functions exhibiting local convexities. This relaxation is possible because some convex utility functions are excluded from the relevant class of utility functions under consideration. Notice that when $\ell=0$, the distributional conditions required for $1+\ell$-ASD coincide with those for AFSD, implying that AFSD is a limiting case of  $1+\ell$-ASD as $\ell$ approaches to 0.  Furthermore, relative to  $1+\ell$-ASD ($0<\ell<1$), ASSD further allows  $\mu_F+\sigma^2_F/2\ge \mu_G+\sigma^2_G/2$ while guaranteeing that $F_T$ is  preferred to $G_T$ for all non-decreasing and concave utility functions. This is due to the exclusion of non-decreasing utility functions with local convexities from the class of utility functions  considered. 

Notice also that, unlike the asymptotic fractional-order stochastic dominance $1+c$-ASD introduced by \cite{2026Xie}, the fractional-order parameter $\ell$ does influence on the equivalent distributional condition of $1+\ell$-ASD. As $\ell$ increases from 0 to 1, the conditions required for $1+\ell$-ASD become progressively less restrictive. Consequently, the proposed  $1+\ell$-ASD overcomes the key limitation  inherent in the $1+c$-ASD, that is, the fractional-order  parameter $c$ fails to influence the equivalent condition  under the assumption of  lognormally distributed returns.

\begin{table}[h]\footnotesize
	\centering
	\caption{Moment conditions for the asymptotic dominance of $F_T$ over $G_T$ under different asymptotic stochastic dominance rules as $T\rightarrow\infty$.}
	\label{Table:ST2}
	\resizebox{\textwidth}{!}{%
	\begin{tabular}{*{111}{l}}
		\hline\hline\noalign{\smallskip}
		Stochastic dominance & Moment conditions & \\ 
		\noalign{\smallskip}\hline\noalign{\smallskip}
		AFSD & $\mu_F>\mu_G$ and $\sigma_F=\sigma_G$ & \cite{Huang2020Comment} \\
		ASSD & $\mu_F+\sigma^2_F/2\ge\mu_G+\sigma^2_G/2$ and $\sigma_F\le\sigma_G$ with at least one inequality being strict & \cite{Huang2020Operational} \\
			$1+c$-ASD ($0<c<1$) & $\mu_F+\sigma^2_F/2>\mu_G+\sigma^2_G/2$ and $\sigma_F\le\sigma_G$ & \cite{2026Xie} \\	
		$1+\ell$-ASD ($0<\ell<1$) & $\mu_F+\sigma^2_F/(2\ell)\ge\mu_G+\sigma^2_G/(2\ell)$ and $\sigma_F\le\sigma_G$ with at least one inequality being strict  & Theorem \ref{thm:AFSD} \\ 
		\noalign{\smallskip}\hline
	\end{tabular}
}
\end{table}

 As shown in Tables \ref{Table:ST} and \ref{Table:ST2}, it is noteworthy that the distributional conditions for AFSD, ASSD and $1+\ell$-ASD coincide exactly with those for FSD, SSD  and $1+\ell$-SD under a finite investment horizon, respectively. Nevertheless, the distributional condition for $1+c$-ASD ($0<c<1$) differs from that for $1+c$-SD under a finite investment horizon. 
 
 \begin{example}
 	Consider two assets, $F_T$ and  $G_T$, following the lognormal distributions with parameters presented in Table \ref{tab1+}, that is, $\mu_F=3$, $\sigma^2_F=0.11$, $\mu_G=3.01$ and $\sigma^2_G=0.1$.

 	Since $\sigma_F\neq \sigma_G$, the two assets cannot be ranked using the AFSD rule. Nevertheless,  because  $\mu_F+\sigma^2_F/2=3.055<\mu_G+\sigma^2_G/2=3.06$ and $\sigma_F>\sigma_G$, they can be ranked under the $1+c$-ASD or SSD rules. More specifically,  $G_T\succsim_{1+c-\textit{ASD}}F_T$ as $T\rightarrow\infty$ for $0<c\le 1$. Hence, a substantial discontinuity arises in the asymptotic dominance relation when  moving from $c=0$ to $c>0$. 
 	
 	It follows from Theorem \ref{thm:AFSD} that the two assets can also be ranked using the $1+\ell$-ASD rule, that is, $$G_T\succsim_{1+\ell-\textit{ASD}}F_T\ \ {\rm as}\ \  T\rightarrow\infty\ \ {\rm for}\ \ 0.5\le \ell\le 1,$$ 
 	which establishes a continuous asymptotic dominance relation from AFSD to ASSD. Therefore, the $1+\ell$-ASD overcomes the limitation of $1+c$-ASD rule, that is, the fractional-order parameter $c$ does not influence the distributional conditions such that a discontinuity arises in the asymptotic dominance relation when  moving from $c=0$ to $c>0$. 
 	
 \end{example}

\section{General asymptotic stochastic dominance of degree $1+\ell$}

In this section, to enhance the tractability, we relax the condition $\sigma_F\le\sigma_G$ associated with $1+\ell$-ASD by imposing an additional constraint on the marginal utility, that is, $0<\inf\limits_{x}\frac{U'(x)}{x^{1/\ell-1}}\le \sup\limits_{x}\frac{U'(x)}{x^{1/\ell-1}}<\infty$. 
This modification yields the following class of utility functions.

\begin{defn}\label{gammaU*}
	For $0< \ell\le 1$, let $\mathcal{U}^*_{1+\ell}$ be the class of functions $U$ such that 
	\begin{equation}\label{eq:gammaU*}
	U'(x)\ge 0,\ \ U'(x)/x^{1/\ell-1}\ \ \text{is non-increasing on}\ \ (0,\infty)\ \ {\rm and}\ \    0<\inf\limits_{x}\frac{U'(x)}{x^{1/\ell-1}}\le \sup\limits_{x}\frac{U'(x)}{x^{1/\ell-1}}<\infty. 
	\end{equation}
\end{defn}

\begin{rem}\label{rem:equ++} For each $0< \ell\le 1$, when $U'>0$,  $U\in \mathcal{U}^*_{1+\ell}$ is equivalent to the condition that 	\begin{equation*} R_{U,2}(x)\ge-\big(\frac{1}{\ell}-1\big)\ \  \  {\rm and}\ \  0<\inf\limits_{x}\frac{U'(x)}{x^{1/\ell-1}}\le \sup\limits_{x}\frac{U'(x)}{x^{1/\ell-1}}<\infty.  
	\end{equation*}
\end{rem}

~

The condition $0<\inf\limits_{x}\frac{U'(x)}{x^{1/\ell-1}}\le \sup\limits_{x}\frac{U'(x)}{x^{1/\ell-1}}<\infty$ is not strict. For example, for any $0<\ell<1$ and $0<\rho<1$, the  utility function $U(x)=\ell x^{1/\ell}+(1+x^{1/\ell})^\rho$ belongs to the class  $ \mathcal{U}^*_{1+\ell}$.


Based on the class $\mathcal{U}^*_{1+\ell}$ of utility functions, we define a variant of $1+\ell$-ASD, referred to as \textit{general asymptotic  stochastic dominance of degree $1+\ell$} (general $1+\ell$-ASD) as follows.

\begin{defn}\label{OAFSD}
	For $0< \ell\le 1$, $F_T$ dominates $G_T$ by general   $1+\ell$-ASD if and only if 
	\begin{equation}\label{eq:OAFSD}
	\liminf_{T\rightarrow\infty}\big(\mathbb{E}_F[U(W_T)]-\mathbb{E}_G[U(W_T)]\big)\ge 0\ \ \text{for all}\ \ U\in \mathcal{U}^*_{1+\ell},
	\end{equation}
	and there exists some $U\in\mathcal{U}^*_{1+\ell}$ for which the inequality is strict. 
\end{defn} 

\begin{rem}
	Note that $\mathcal{U}^*_{1+\ell}\subset \mathcal{U}_{1+\ell}$. This implies that for a fixed $0\le \ell\le 1$, $1+\ell$-ASD is a  stronger rule than general  $1+\ell$-ASD, and thus $1+\ell$-ASD implies general  $1+\ell$-ASD if there exists some $U\in\mathcal{U}^*_{1+\ell}$ for which the inequality \eqref{eq:OAFSD} is strict. 
\end{rem}



In the following theorem, we provide the equivalent conditions for general $1+\ell$-ASD, $0< \ell< 1$, under the assumption of lognormally distributed returns. 

\begin{thm}\label{thm:OAFSD}
	Suppose that $F_T$ and $G_T$ follow lognormal distributions with means  $e^{T(\mu_F+\sigma_F^2/2)}$ and $e^{T(\mu_G+\sigma_G^2/2)}$, respectively. For $0< \ell\le 1$ and as  $T\rightarrow\infty$, the distribution $F_T$ dominates $G_T$ by general $1+\ell$-ASD, if and only if  
	\begin{equation}
		\label{ocondition1}
		\mu_F+\frac{\sigma^2_F}{2\ell}> \mu_G+\frac{\sigma^2_G}{2\ell},\ \ \text{or}\ \ \mu_F+\frac{\sigma^2_F}{2\ell}= \mu_G+\frac{\sigma^2_G}{2\ell}\ \ \text{with}\ \ \sigma_F< \sigma_G. 
	\end{equation}
\end{thm}

The following example illustrates the advantage of general $1+\ell$-ASD in comparing assets over a sufficiently long investment horizon. 

\begin{example}
	Consider an agent with the utility function $U$ satisfying
	 \begin{equation}\label{DUM++}U'(t)	=\left\{\begin{array}{ll}		t^{1/\ell-1},  & \ \ t\le 1,\\	
	 \big(b+(1-b)(t-2)^2\big)t^{1/\ell-1},  & \ \ 1< t\le 2,\\	
	 		b t^{1/\ell-1},  & \ \ t>2,	\end{array}\right.\end{equation}
where $0<b<1$ and $0<\ell<1$ are constants. It is obvious that the utility function $U$ is not concave and, therefore, cannot be characterized by ASSD rule as $T\rightarrow \infty$. 
	
	Notice easily that $U'(x)\ge 0$, $b\le U'(x)/x^{1/\ell-1}\le 1$ for all $x> 0$ and $U'(x)/x^{1/\ell-1}$, $x> 0$ is non-increasing. Hence, the utility function $U$ of the form given in \eqref{DUM++} satisfies the condition \eqref{eq:gammaU*}, implying that  $U\in\mathcal{U}^*_{1+\ell}$. 


	Assume that two assets, $F_T$ and  $G_T$, follow  lognormal distributions with means $e^{1.31125T}$ and $e^{1.2898T}$, respectively, where $\mu_F=1.30$, $\sigma_F=0.15$, $\mu_G=1.28$ and $\sigma_G=0.14$. 
	
	Since $\sigma_F\neq \sigma_G$, the two assets cannot be ranked using the AFSD rule. Moreover, because  $\mu_F+\sigma^2_F/(2\ell)>\mu_G+\sigma^2_G/(2\ell)$ holds  for all $0<\ell\le 1$ while $\sigma_F>\sigma_G$, they also cannot be ranked under the $1+\ell$-ASD or  ASSD rules. However, Theorem \ref{thm:OAFSD}  implies that $F_T$ dominates $G_T$ by the general $1+\ell$-ASD for $0< \ell< 1$. Consequently, the agent with the aforementioned  utility function $U$ of type \eqref{DUM++}  would prefer asset $F_T$ over $G_T$ as $T\rightarrow\infty$.  
\end{example}

\begin{rem}\label{gammaMCSD+}

In the limiting case $\ell=0$, $\mathcal{U}^*_{1}$ is the class of utility functions $U$ such that $U'(x)>0$ and $\lim\limits_{\ell\rightarrow0}(U'(x)/x^{1/\ell-1})$ is non-increasing, strictly positive and  bounded on $(0,+\infty)$. Consequently,  $\mathcal{U}^*_{1}$ is a subset of $\mathcal{U}_{1}$. In this case, the condition \eqref{ocondition1} is equivalent to either  $\mu_F>\mu_G$ with $\sigma_F=\sigma_G$ or  $\sigma_F>\sigma_G$. This implies that the strict variance condition $\sigma_F=\sigma_G$ required for AFSD is relaxed by imposing an additional constraint on the marginal utility, that is, $\lim\limits_{\ell\rightarrow0}(U'(x)/x^{1/\ell-1})$ is strictly positive and bounded on $(0,+\infty)$. 

In the case $\ell=1$, $\mathcal{U}^*_{2}$ is a subset of $\mathcal{U}_{2}$ obtained by imposing an additional constraint on the marginal utility, that is,  $0<\inf\limits_{x}U'(x)\le \sup\limits_{x}U'(x)<\infty$. In this case, the condition \eqref{ocondition1} corresponds to the equivalent distributional condition for $F_T$ dominating $G_T$ by operational ASSD, as established by \cite{Huang2020Operational} for DMs whose preferences satisfy $U'(x)>0$, $U''(x)\le 0$ and $0<\inf\limits_{x}U'(x)\le \sup\limits_{x}U'(x)<\infty$. These conditions are summarized in Table \ref{Table:ST3}. This indicates that the operational ASSD is a limit case ($\ell=1$) of general $1+\ell$-ASD. Thus, the general  $1+\ell$-ASD implies the operational ASSD. 

Table \ref{Table:ST3} shows that the moment conditions of operational 
$1+c$-ASD introduced by \cite{2026Xie} are identical to those for operational AFSD. Nevertheless, the preferences of the two rules are addressed from
distinct perspectives. The operational AFSD focuses on  preferences characterized by
strictly positive and bounded marginal utilities. In contrast,
the operational 
$1+c$-ASD captures the preferences of DMs
whose utility functions exhibit local convexities. It is also worth noting that, unlike the operational 
$1+c$-ASD, the fractional-order parameter $\ell$ does   influence the equivalent condition of general $1+\ell$-ASD under the lognormal distribution assumption. As $\ell$ increases from 0 to 1, the conditions required for general $1+\ell$-ASD become progressively less restrictive.

\begin{table}[H]\footnotesize
	\centering
	\caption{Moment conditions for the asymptotic dominance of $F_T$ over $G_T$ under variants of  asymptotic stochastic dominance rules as $T\rightarrow\infty$.}
	\label{Table:ST3}
	\resizebox{\textwidth}{!}{%
	\begin{tabular}{*{111}{l}}
		\hline\hline\noalign{\smallskip}
		Stochastic dominance & Moment conditions & Extra requirements & \\ 
		\noalign{\smallskip}\hline\noalign{\smallskip}
	Operational AFSD & $\mu_F+\sigma^2_F/2>\mu_G+\sigma^2_G/2$ & $0<\inf\limits_{x}U'(x)\le \sup\limits_{x}U'(x)<\infty$ & \cite{Huang2020Operational} \\
		Operational	ASSD & $\mu_F+\sigma^2_F/2>\mu_G+\sigma^2_G/2$ or  & $0<\inf\limits_{x}U'(x)\le \sup\limits_{x}U'(x)<\infty$ & \cite{Huang2020Operational} \\
	 & $\mu_F+\sigma^2_F/2=\mu_G+\sigma^2_G/2$ with $\mu_F>\mu_G$ &   &  \\				
	Operational	$1+c$-ASD ($0<c<1$) & $\mu_F+\sigma^2_F/2>\mu_G+\sigma^2_G/2$ & $U'(0)<\infty$  & \cite{2026Xie} \\	
	General	$1+\ell$-ASD ($0<\ell<1$) & $\mu_F+\sigma^2_F/(2\ell)>\mu_G+\sigma^2_G/(2\ell)$ or & $0<\inf\limits_{x}\frac{U'(x)}{x^{1/\ell-1}}\le \sup\limits_{x}\frac{U'(x)}{x^{1/\ell-1}}<\infty$ & Theorem \ref{thm:OAFSD} \\
		& $\mu_F+\sigma^2_F/(2\ell)=\mu_G+\sigma^2_G/(2\ell) $ with $\sigma_F<\sigma_G$  & &  \\ 
		\noalign{\smallskip}\hline
	\end{tabular} 
}
\end{table}

\end{rem}

\section{Empirical analysis for asset selection in long-term investments}

This section applies the proposed asymptotic fractional-order  stochastic dominance rules with bounded relative risk aversion to the ranking of  multiple assets in long-term investment decisions, thereby illustrating  the practical usefulness of these newly proposed rules in optimal asset selection over a sufficiently long investment horizon. The empirical analysis is conducted by considering five representative asset classes:  gold futures, silver futures, Brent crude oil futures,10-year U.S. Treasury bonds and stocks (NASDAQ Composite Index). The applications of Theorems \ref{thm:AFSD} and \ref{thm:OAFSD} are illustrated through a three-step procedure. In Step I, under the assumption that the daily   returns of
these assets follow lognormal distributions,  
the empirical
parameters of daily log-returns for each asset are estimated using the real data. In  Step II, we study the estimation of the fractional-order parameter $\ell$ in the $1+\ell$-ASD rule.  Finally, in Step III, based on the estimated index $\ell$ and the empirical
parameters of daily log-returns for each asset, we conduct an analysis to the ranking of all five assets under consideration by the newly proposed rules, thereby elucidating the practical application of the $1+\ell$-ASD and general   $1+\ell$-ASD rules in optimal asset selection over a sufficiently long investment horizon. 

	\subsection{Estimation of daily log-return parameters}

We compare the log-returns and daily changes of the following five assets: ($A_1$) gold futures, ($A_2$) silver futures, ($A_3$) Brent crude oil futures, ($A_4$) the 10-year U.S. Treasury bonds, and ($A_5$) the stocks in NASDAQ Composite Index. 
Let $R_{i,t}$ denote the observed daily level of asset $A_i$, $i\in\{1,\dots,5\}$. Then, the log-return $r_{i,t}$ of the $i$-th asset on day $t$ is calculated as 
\begin{equation}
	r_{i,t} \;=\; \log\!\left(\frac{R_{i,t}}{R_{i,t-1}}\right),
	\quad t=2,\dots,T_i,
\end{equation}
where $T_i$ is the observed days for the $i$-th asset. 

In practice, it is often necessary to identify the estimations based on daily observations $r_{i,2}$, $\ldots$, $r_{i,T_i}$. Assume that for each $i\in\{1,\ldots,5\}$, the observations $r_{i,t}$,  $t=2,\ldots,T_i$ follow a normal distribution $N(\mu_i,\sigma^2_i)$. 
Let $F_{i,T_i}$ denote the empirical CDF  
of $r_{i,2}$, $\ldots$, $r_{i,T_i}$. We further assume that $F_{i,T_i}\rightarrow N(\mu_i,\sigma^2_i)$ in probability as $T_i\rightarrow \infty$. Evidently, this is the
minimum requirement for any meaningful approximation of the true
distributions by using their empirical ones. Hence, the mean  and volatility parameters of daily log-return for the $i$-th asset are estimated by the sample moments
\begin{equation}\label{estimated}
	\hat\mu_i \;=\; \frac{1}{n_i}\sum_{t=2}^{T_i} r_{i,t}
	\ \ \ {\rm and}\ \ \ 
	\hat\sigma_i \;=\; \sqrt{\frac{1}{n_i-1}\sum_{t=2}^{T_i}\bigl(r_{i,t}-\hat\mu_i\bigr)^2},
\end{equation}
where $n_i=T_i-1$ is the number of observed daily log-returns for the $i$-th asset.

Using daily observation data for the period from 2016 to 2025\footnote{The data is downloaded from https://www.investing.com/}, we obtain the empirical 
parameters of the log-return for each asset, as reported in Table \ref{tab1}. 

\begin{table}[h]
	\centering
	\caption{Estimated parameters of the daily  log-returns for all five assets (Ten-year history).}
	\label{tab1}
	\begin{tabular}{lcccc}
		\toprule
		Asset $A_i$ & $n_i$ & $\hat\mu_i$ & $\hat\sigma_i$ & $\hat\mu_i+\hat\sigma_i^2/2$ \\
		\midrule
		($A_1$)	Gold futures        & 2182 & 0.000644 & 0.010355 & 0.000698 \\
		($A_2$)	Silver futures      & 2063 & 0.000838 & 0.020720 & 0.001052 \\
		($A_3$)	Brent crude oil futures   & 2581 & 0.000190 & 0.024369 & 0.000487 \\
		($A_4$)	10-year U.S. Treasury bonds   & 2604 & 0.000237 & 0.029493 & 0.000672 \\
		($A_5$)	Stocks in NASDAQ Composite Index   & 2513 & 0.000619 & 0.013910 & 0.000716 \\
		\bottomrule
	\end{tabular}
\end{table}

	\subsection{Estimation of the fractional-order parameter $\ell$ in  $1+\ell$-ASD}
	
This subsection addresses the estimation of the fractional-order parameter $\ell$ in the $1+\ell$-ASD rule. 
Since the asymptotic dominance relation of $1+\ell$-ASD becomes stronger as $\ell$ decreases, it is natural to determine the smallest value (or infimum) of $\ell_{ij}$ for which $A_i\succeq_{1+\ell_{ij}-\textit{ASD}}A_j$. Denote this value by $\ell^{\min}_{ij}$,
assuming that it exists (otherwise, we take the infimum). 



From Theorem \ref{thm:AFSD}, it follows that $A_i\succeq_{1+\ell_{ij}-\textit{ASD}}A_j$ if and only if $\mu_i+\frac{1}{2\ell_{ij}}\sigma^2_i\ge \mu_j+\frac{1}{2\ell_{ij}}\sigma^2_j$ and 
$\sigma_i\leq \sigma_j$. 
Notice that if $\mu_i< \mu_j$ and $\sigma_i \leq \sigma_j$, then there exists no $\ell_{ij}\in(0,1]$ satisfying the inequality $\ell_{ij}\geq \frac{\sigma_j^2-\sigma_i^2}{2(\mu_i-\mu_j)}$, and thus $A_i\nsucceq_{1+\ell_{ij}-\textit{ASD}}A_j$. Consequently, we require at least $\mu_i\geq \mu_j$ and $\sigma_i\leq \sigma_j$. This implies that $\ell_{ij}^{\min}= \frac{\sigma_j^2-\sigma_i^2}{2(\mu_i-\mu_j)}$. Notice that under the consistency of the sample mean and variance estimators, if $\mu_i>\mu_j$,  $\sigma_i<\sigma_j$, and $0<\ell_{ij}^{\min}<1$, then by the Continuous Mapping Theorem, we have that $\frac{\hat\sigma_i^{\,2}-\hat\sigma_j^{\,2}}{2(\hat\mu_j-\hat\mu_i)}\rightarrow \ell_{ij}^{\min}=\frac{\sigma_j^2-\sigma_i^2}{2(\mu_i-\mu_j)}$, where $\hat\mu_i$ and $\hat\sigma_i$ are given in \eqref{estimated}. Based on the above observation, we propose the following estimator $\hat{\ell}_{ij}^{\min}$ as an
estimation of $\ell^{\min}_{ij}$ between two assets $A_i$ and $A_j$:
	\begin{equation}
	\hat{\ell}_{ij}^{\min}
		\;=\;
		\min\!\left\{
		1,\;
		\frac{\hat\sigma_i^{\,2}-\hat\sigma_j^{\,2}}{2(\hat\mu_j-\hat\mu_i)}
		\right\}.
		\label{eq:chat}
	\end{equation}
 

Using the estimator defined in \eqref{eq:chat}, we obtain the estimated fractional-order index $\hat \ell_{ij}^{\min}$ for each pair of assets $A_i$ and $A_j$, $i,j=1,\ldots,5$. These estimates are reported in Table \ref{tab2}. From the numerical results presented in Table \ref{tab2}, the asymptotic dominance relation between any two
assets $A_i$ and $A_j$, $i,j=1,\ldots,5$ under the $1+\ell$-ASD rule is determined.

	\begin{table}[h]
	{\centering
		\caption{Pairwise estimated values of $\hat \ell_{ij}^{\min}$.}
		\label{tab2}
		\resizebox{\textwidth}{!}{%
			\begin{tabular}{lccccc}
				\toprule
				& ($A_1$) Gold futures
				& ($A_2$) Silver futures
				& ($A_3$) Brent crude oil futures
				& ($A_4$) U.S. Treasury bonds
				& ($A_5$) Stocks\\
				\midrule
				($A_1$) Gold futures
				& --        & -- &  0.535923 &  0.936849 &  -- \\
				($A_2$) Silver futures
				& -- & --        &  0.126958 &  0.366464 & -- \\
				($A_3$) Brent crude oil futures
				&  -- &  -- & --        & -- &  -- \\
				($A_4$) U.S. Treasury bonds
				&  -- &  -- & -- & --        &  -- \\
				($A_5$) Stocks
				&  -- & -- &  0.466620 &  0.885265 & --        \\
				\bottomrule
			\end{tabular}%
	} }
	Notes: These values of $\hat \ell_{ij}^{\min}$ are estimated by Eq \eqref{eq:chat} and numerical results in Table \ref{tab1}. The notation ``--'' indicates that there exists no $0<\ell\le 1$ such that $A_i\succsim_{1+\ell-\textit{ASD}}A_j$.
\end{table}


\subsection{Application of the $1+\ell$-ASD to the optimal asset selection in long-term investment decisions}
	
This subsection presents an empirical analysis illustrating the application of the $1+\ell$-ASD and the general $1+\ell$-ASD rules to the ranking of five representative asset classes $A_i$, $i=1,\ldots,5$ in long-term investment decisions. The  empirical analysis is divided into two parts. The first part discusses the pairwise comparisons 
of individual assets to show the applicability of $1+\ell$-ASD and general $1+\ell$-ASD rules. The second part extends  
the analysis to a simultaneous comparison of all five  
assets under consideration, thereby elucidating the 
practical application of the general $1+\ell$-ASD 
rule in optimal asset selection over a sufficiently long investment 
horizon.

\textit{Part I}. Based on the empirical parameters of the daily log-returns for all five assets presented in Table \ref{tab1}, we observe that the AFSD rule fails to rank any pair of assets over a long investment horizon since $\sigma_i=\sigma_j$, $i\neq j$ does not hold for any $i,j\in\{1,\ldots,5\}$.

The numerical results presented in Table \ref{tab2} reveal the asymptotic dominance relation between any two
 assets $A_i$ and $A_j$, $i,j=1,\ldots,5$ under the $1+\ell$-ASD rule. 
 
 \begin{itemize}
 	\item Gold futures are preferred to 10-year U.S. Treasury bonds in long-term investments by all DMs whose preferences are monotone under the $1+\ell$-ASD rule for  $0.936849\le\ell\le 1$. Similarly, gold futures are preferred to Brent crude oil futures under the $1+\ell$-ASD rule for  $0.535923\le\ell\le 1$. Consequently, for the asset classes of gold futures, Brent crude oil futures, and 10-year U.S. Treasury bonds in long-term investments, gold futures are preferred by all DMs whose preferences are monotone under the $1+\ell$-ASD rule for  $0.936849\le\ell\le 1$.
 	\item Silver futures are preferred to Brent crude oil futures in long-term investments  under the $1+\ell$-ASD rule for  $0.126958\le\ell\le 1$. Also, silver futures are preferred to 10-year U.S. Treasury bonds  under the $1+\ell$-ASD rule for  $0.366464\le\ell\le 1$.
 	\item Stocks in NASDAQ Composite Index are preferred to Brent crude oil futures in long-term investments under the $1+\ell$-ASD rule for  $0.466620\le\ell\le 1$. Additionally, stocks in NASDAQ Composite Index are preferred to 10-year U.S. Treasury bonds under the same rule for  $0.885265\le\ell\le 1$.
 	\item For the remaining pairs of assets in long-term investments, no asymptotic dominance relations exist under the $1+\ell$-ASD rule for any $0<\ell \le 1$.  
 \end{itemize}
 
 The above analysis shows that the $1+\ell$-ASD rule can rank certain assets in long-term investments that the AFSD rule fails to distinguish. Nevertheless, there remain some assets for which $1+\ell$-ASD rule still does not yield a clear comparison. For example, for gold futures $A_1$ and silver futures $A_2$, based on the numerical
 results reported in Table \ref{tab2}, 
 we have that
 $$\mu_1+\frac{\sigma^2_1}{2\ell}<\mu_2+\frac{\sigma^2_2}{2\ell}\ \ {\rm for\ all}\ \ 0<\ell\le 1 \ \ {\rm and} \ \ \sigma_1=0.010355<0.020720=\sigma_2.$$
Consequently, it follows from Theorem \ref{thm:AFSD} that gold futures and silver futures cannot be ranked under
$1+\ell$-ASD rule in long-term investments for any $0<\ell\le 1$. However, because $\mu_2+\frac{\sigma^2_2}{2\ell}>\mu_1+\frac{\sigma^2_1}{2\ell}$ holds for  all $0<\ell\le 1$, Theorem \ref{thm:OAFSD} implies that all DMs
whose utility functions satisfy Condition \eqref{eq:gammaU*} with $0<\ell\le 1$ would prefer silver futures ($A_2$) over gold futures ($A_1$) under general 
$1 + \ell$-ASD rule. This result
highlights the practical advantage of the general 
$1 + \ell$-ASD rule in asset selection over a long investment horizon, particularly in cases where the
$1 + \ell$-ASD rule fails to provide a clear ranking of 
assets under consideration.

\textit{Part II}. Consider a scenario in which DMs make an
investment choice among multiple assets $A_i$, $i=1,\ldots,5$  over a sufficiently
long investment horizon. The discussion in Part I implies that the $1+\ell$-ASD rule is likewise inapplicable in
this context for all $0< \ell\le 1$.
In contrast, the general $1+\ell$-ASD offers a practicable rule for ranking these assets.

Under the assumption that the daily log-returns of these
assets follow normal distributions,  Theorem  \ref{thm:OAFSD} implies
that DMs would prefer the asset that attains either the
highest value of $\mu +\sigma^2/(2\ell)$ or, among those with the same value of $\mu +\sigma^2/(2\ell)$, the smallest $\sigma$. The numerical results in Table \ref{tab2} indicate that there exists no $\ell\in(0,1]$ for which  $\mu_i +\sigma_i^2/(2\ell)$, $i=1,\ldots,5$ are identical.  Nevertheless, for all $\ell\in(0,0.366464)$, 10-year U.S. Treasury bonds attain the
highest value of $\mu +\sigma^2/(2\ell)$, whereas for $\ell\in(0.366464,1]$, the  silver futures attain the
highest such value. This result implies that all
DMs whose utility functions $U$ exhibit bounded relative risk aversion and satisfy $0<U'(x)/x^{1/\ell-1}<+\infty$ would optimally invest
in 10-year U.S. Treasury bonds for $\ell\in(0,0.366464)$ and in silver futures for $\ell\in(0.366464,1]$ over very long-term investment horizons.

\section{Conclusion}

This paper introduces an alternative of fractional-order stochastic dominance rule, that is, the $(1+\ell)$th-order stochastic
dominance ($1+\ell$-SD). It captures the consensus in the distribution
ranking of individuals whose 
relative risk aversion shares a common lower
bound, that is, $-(1/\ell-1)$. This rule  complements the $(1+c)$th-order  stochastic dominance rule established by \cite{Huang2020Fractional}, which imposes a lower bound on the 
absolute risk aversion, as well as the $(1+\gamma)$th-order stochastic
dominance rule developed by \cite{Mullerd2017Between} through the requirement of a boundary condition on the ratio of marginal utilities. The equivalent distributional condition for $1+\ell$-SD is provided. 

Furthermore, a novel asymptotic fractional-order stochastic dominance, referred to as asymptotic $(1+\ell)$th-order stochastic
dominance ($1+\ell$-ASD), is proposed by incorporating the influence of
the investment horizon into the framework of $1+\ell$-SD. Relative to the existing asymptotic fractional-order rule \citep{2026Xie}, the strength of our approach lies in overcoming the key shortcoming that the fractional-order parameter does not influence the distributional conditions. Owing to this advantage, the conditions required for  $1+\ell$-ASD become progressively less restrictive as $\ell$ increases from 0 to 1. Notably, under the assumption that returns are lognormally distributed, we establish equivalent conditions for  $1+\ell$-ASD rule without imposing the non-negativity constraint on the mean of log-return, a restriction usually required by the existing asymptotic stochastic dominance rules.  Similar to AFSD and ASSD rules,  the conditions for $1+\ell$-ASD, $0<\ell<1$ coincide with those for $1+\ell$-SD under a finite horizon. To enhance the tractability of $1+\ell$-ASD, under an  additional assumption that marginal utilities of DMs satisfy $0<\inf\limits_{x}\frac{U'(x)}{x^{1/\ell-1}}\le \sup\limits_{x}\frac{U'(x)}{x^{1/\ell-1}}<\infty$, we propose a variant of $1+\ell$-ASD, referred to as general $1+\ell$-ASD. We also provide the equivalent distributional conditions for general $1+\ell$-ASD,  which relax the volatility constraint inherent in $1+\ell$-ASD.  Empirical results illustrate that our distributional conditions are practically significant for DMs with bounded relative risk aversion. Because stochastic dominance rules are applicable to a broad range of  long-term investment decision contexts beyond those illustrated
in this paper, we hope that the proposed rules of $1+\ell$-ASD and general $1+\ell$-ASD can contribute to the further application of the
stochastic dominance methodology in the optimal long-term portfolio selection.

Our study focuses on the asymptotic fractional-order stochastic dominance between AFSD and ASSD, and relies  on  the assumption of lognormally distributed returns, consistent  with the existing research such as \cite{Levy2016Aging,Levy2020Aging} and \cite{2026Xie}. Future research directions include developing asymptotic  stochastic dominance rules that bridge all adjacent integer degrees, as well as extending the asymptotic fractional-order stochastic dominance rules to more general return distributions.

\subsection*{Declaration of competing interest} The authors declare that they have no known competing financial interests or personal relationships that could have
appeared to influence the work reported in this paper.

\subsection*{Acknowledgements}  The research was supported by the National Natural Science Foundation of China (Grants
No. 72271113) and the Natural Science Foundation of Jiangxi Province (Grants No. 20232ACB201003). All authors share first authorship.

\section*{Appendix}
\appendix
\section{Preliminary lemmas}\label{Appendix:B}

To derive the equivalent distributional conditions for 1+$\ell$-SD, we need  the following lemma, which is verified in the proof of Theorem 2.4 in \cite{Mullerd2017Between}.

\begin{lem}\label{lem:1}
	If $\int_0^x\beta(t){\rm d}t\ge 0$ for all $x\ge 0$, then $\int_0^{+\infty}\alpha(x)\beta(x){\rm d}x\ge 0$ for all $\alpha(x)\ge0$ that is non-increasing. 
\end{lem}

Let $Z_{p}$ denote the quantile of order $p$ of a normal distribution $N(0,1)$, and let $Z_{\Lambda_p}$ denote the quantile of order $p$ of the lognormal distribution $\Lambda\big(e^{\mu+\sigma^2/2}, e^{2\mu+\sigma^2}(e^{\sigma^2}-1)\big)$, respectively. 
It then follows that $Z_{\Lambda_p}=e^{\mu+Z_{p}\sigma}$. Therefore, for the CDFs $F_T$ and $G_T$, we have that 
$$Z_{\Lambda_p}(F_T)=e^{T\mu_F+Z_{p}\sqrt{T}\sigma_F}\ \ {\rm and}\ \ Z_{\Lambda_p}(G_T)=e^{T\mu_G+Z_{p}\sqrt{T}\sigma_G} .$$

The following lemma presents a property of $Z_{p}$ and $Z_{\Lambda_p}$ that will be useful in deriving the equivalent distributional conditions under which $F_T$ dominates $G_T$ according to $1+\ell$-SD rule. Since the result is straightforward, we omit the proof.

\begin{lem}\label{Huanglemma}
	Assume that $F_T$ and $G_T$ are two lognormal distributions with means  $e^{T(\mu_F+\sigma_F^2/2)}$ and $e^{T(\mu_G+\sigma_G^2/2)}$ respectively, where $\sigma_F\neq \sigma_G$. 
	Let $Z_{p_0}=\sqrt{T}(\mu_F-\mu_G)/(\sigma_G-\sigma_F)$ such that  \begin{equation}\label{equ}
	e^{T\mu_F+Z_{p_0}\sqrt{T}\sigma_F}=e^{T\mu_G+Z_{p_0}\sqrt{T}\sigma_G}=e^{T(\frac{\mu_F}{\sigma_F}-\frac{\mu_G}{\sigma_G})/(\frac{1}{\sigma_F}-\frac{1}{\sigma_G})}=Z_{\Lambda_{p_0}}.
	\end{equation} 
	Then, $Z_{\Lambda_{p_0}}$ is the intersection point of the CDFs $F_T$ and $G_T$. 
\end{lem}

\section{Proofs}\label{Appendix:A}

\subsection{Proof of Theorem \ref{Cthm}}

	($a$). {\bf Sufficiency}.  Suppose that the inequality \eqref{cfracSD} holds. Denote $\Lambda_{\ell}(x)=\int^x_{0}\big(G(t)-F(t)\big){\rm d}h_{\ell}(t)$.
	
	Note that 	
	\begin{equation}\label{eq:f1}
		\mathbb{E}[U(X)]-\mathbb{E}[U(Y)]=\int_{0}^{+\infty}\big(G(t)-F(t) \big)U'(t) {\rm d}t
		=\ell\int_{0}^{+\infty}\frac{U'(t)}{t^{1/\ell-1}}\big(G(t)-F(t)\big) {\rm d}h_{\ell}(t).
	\end{equation}
	Since for any $U\in\mathcal{U}_{1+\ell}$, the function $\frac{U'(x)}{x^{1/\ell-1}}$ is non-increasing and non-negative on $(0,\infty)$, it follows from Lemma \ref{lem:1} that   $\int_{0}^{\infty}\frac{U'(t)}{t^{1/\ell-1}}\big(G(t)-F(t) \big) {\rm d}h_{\ell}(t)\ge 0$, and thus  $\mathbb{E}[U(X)]\ge\mathbb{E}[U(Y)]$ by \eqref{eq:f1}. 
	
	It remains to verify that there exists some  \(U\in\mathcal U_{1+\ell}\) such that $\mathbb{E}[U(X)]>\mathbb{E}[U(Y)]$. Suppose that the inequality \eqref{cfracSD} is strict for $x=x_0$. Let $U_0$ be a utility function with
	$$
U_0'(x)=
	\begin{cases}
		h_\ell'(x), & \text{if }0<x<x_0,\\
		0, & \text{if }x>x_0.
	\end{cases}
	$$
	It is easy to verify that $U_0\in\mathcal{U}_{1+\ell}$. Moreover,
	$$\mathbb{E}[U_0(X)]-\mathbb{E}[U_0(Y)]=\int_0^{x_0}\bigl(G(t)-F(t)\bigr){\rm d}h_\ell(t)>0.$$
	
	($b$). {\bf Necessity}. It can be proved by contradiction.
	If there exists $x_0>0$ such that the inequality  \eqref{cfracSD} does not hold, that is, $\int^{x_0}_{0}\big(G(t)-F(t)\big){\rm d}h_{\ell}(t)=\Lambda_{\ell}(x_0)<0$. It follows from the continuity that there exists a interval $x_0\in (x_1,x_2)\subset(0,+\infty)$ such that $\Lambda_{\ell}(x)<0$ for all $x\in(x_1,x_2)$. Let $g$ be a strictly decreasing and twice continuously differentiable function satisfying $g(x_1)=1$ and $g(x_2)=0$. Assume that $U_0$ is a utility function with 
	\begin{equation*} U'_0(x)
		=\left\{\begin{array}{ll}
			x^{1/\ell-1},  & \ \text{if}\ x\le x_1,\\
			g(x)x^{1/\ell-1},  & \ \text{if}\ x_1< x< x_2,\\
			0,  & \ \text{if}\   x\ge x_2.\\
		\end{array}\right.
	\end{equation*}
	It is easy to verify that $U_0\in \mathcal{U}_{1+\ell}$, and 
	\begin{equation*} \kappa_{\ell}(x):=-U''_0(x)+\frac{U'_0(x)(1/\ell-1)}{x}=
		\left\{\begin{array}{ll}
			0,  & \ \text{if}\ 0\le x\le x_1,\\
			-g'(x)x^{1/\ell-1},  & \ \text{if}\ x_1< x< x_2,\\
			0,  & \ \text{if}\   x\ge x_2 \\
		\end{array}\right. 
	\end{equation*}
	is strictly positive if and only if $x\in(x_1,x_2)$. Integrating \eqref{eq:f1} by parts, it yields that 
	\begin{eqnarray}\label{eq:omaintm3}
		\mathbb{E}[U_0(X)]-\mathbb{E}[U_0(Y)]&=& U'_0(+\infty)\frac{\Lambda_{\ell}(+\infty)}{h'_{\ell}(+\infty)}+\int_{0}^{+\infty} \frac{\Lambda_{\ell}(x)}{h'_{\ell}(x)}\big(-U''_0(x)+\frac{U'_0(x)(1/{\ell}-1)}{x}\big){\rm d} x\nonumber \\
		&= & \int_{x_1}^{x_2} \frac{\Lambda_{\ell}(x)}{h'_{\ell}(x)}\kappa_{\ell}(x){\rm d}x<0,
	\end{eqnarray}
	which is a contradiction to  $\mathbb{E}[U_0(X)]\ge \mathbb{E}[U_0(Y)]$ for all $U\in \mathcal{U}_{1+{\ell}}$. 
	
Finally, we prove by contradiction  that there exists some $x^*>0$ such that $\int_0^{x^*}\bigl(G(t)-F(t)\bigr){\rm d}h_\ell(t)>0$.  Indeed, if 
	$$\int_0^x\bigl(G(t)-F(t)\bigr){\rm d}h_\ell(t)=0 \text{ for all }x>0,$$
	then $F$ and $G$ must be identical. Consequently, there exists no utility function $U$ for which \(\mathbb E_F[U(X)]>\mathbb E_G[U(Y)])\). 

\subsection{Proof of Proposition \ref{prop-Cthm}}

($a$). {\bf Necessity}. 	Taking $x=\infty$ in Inequality \eqref{cfracSD} shows that a necessary condition for  $X\succsim_{1+\ell-\textit{SD}}Y$ is $\int_0^{\infty}\big(G(t)-F(t)\big){\rm d} h_{\ell}(t)\ge0$, which yields 
	$$\int_0^{x_0}\bigl(G(t)-F(t)\bigr){\rm d}h_\ell(t)\geq\int_{x_0}^{+\infty}\bigl(F(t)-G(t)\bigr){\rm d}h_\ell(t).$$
	Moreover, since $G(t)-F(t)\geq 0$ for $t\leq x_0$, if $\int_0^{x_0}\bigl(G(t)-F(t)\bigr){\rm d}h_\ell(t)=0$,
	then
	$\int_0^x\bigl(G(t)-F(t)\bigr){\rm d}h_\ell(t)=0$   for all $x\leq x_0$. 
	For $x>x_0$, since $G(t)-F(t)\leq 0$ for $t\geq x_0$, it follows that  $\int_0^x\bigl(G(t)-F(t)\bigr){\rm d}h_\ell(t)=\int_{x_0}^x\bigl(G(t)-F(t)\bigr){\rm d}h_\ell(t)\leq 0$. However, Inequality \eqref{cfracSD} implies $\int_0^x\bigl(G(t)-F(t)\bigr){\rm d}h_\ell(t)\ge0$. 
	Thus,   $\int_0^x\bigl(G(t)-F(t)\bigr){\rm d}h_\ell(t)=0$ for all $x>0$, which contradicts the strict condition in Theorem \ref{Cthm}. Consequently, it follows that $\int_0^{x_0}\bigl(G(t)-F(t)\bigr){\rm d}h_\ell(t)>0$ must hold.

($b$). {\bf Sufficiency}. 	To establish the sufficiency, we distinguish between two cases.  For $x\le x_0$, it is obvious that $\int_{0}^{x}\big(G(t)-F(t) \big){\rm d}h_{\ell}(t)\ge 0$. For $x> x_0$, we have that $\int_{0}^{x}\big(G(t)-F(t) \big){\rm d}h_{\ell}(t)=\int_{0}^{\infty}\big(G(t)-F(t) \big){\rm d}h_{\ell}(t)-\int_{x}^{\infty}\big(G(t)-F(t) \big){\rm d}h_{\ell}(t)\ge 0$. 
Moreover, $\int_{0}^{x_0}\big(G(t)-F(t) \big){\rm d}h_{\ell}(t)>0$, and thus  $X\succeq_{1+\ell\text{-}SD}Y$.

\subsection{Proof of Theorem \ref{fractional}}

	($a$). {\bf Sufficiency}.  
If $\sigma_F=\sigma_G$, then  $\mu_F>\mu_G$. It follows from Theorem 4 of \cite{1973Dominance} that $F_T$ dominates $G_T$ by FSD, 
that is,  $F_T\succsim_{1-\textit{SD}}G_T$, and thus $F_T\succsim_{1+\ell-\textit{SD}}G_T$ for all $0< \ell\le 1$ by the hierarchy property of $1+\ell$-SD and the strict inequality  $\mathbb{E}_F[U(W_T)]>\mathbb{E}_G[U(W_T)]$ for $U(x)=\log x\in \mathcal{U}_{1+\ell}$, $0<\ell\le 1$. We now consider the other case where  $\mu_F\ge \mu_G$ and $\sigma_F< \sigma_G$. In this case, there exists a non-negative value  $Z_{p_0}=\sqrt{T}(\mu_F-\mu_G)/(\sigma_G-\sigma_F)$ such that  the equality \eqref{equ} holds. For any value $Z_{p}>Z_{p_0}$, we have that $Z_{\Lambda_p}(F_T)<Z_{\Lambda_p}(G_T)$ and $F_T$ is above  $G_T$. Conversely, for any value $Z_{p}<Z_{p_0}$, we have that $Z_{\Lambda_p}(F_T)>Z_{\Lambda_p}(G_T)$ and $F_T$ is below  $G_T$. Thus, $F_T$ and $G_T$ are single crossing at $Z_{\Lambda_{p_0}}$  and  $\int_{0}^{Z_{\Lambda_{p_0}}}\big(G_T(t)-F_T(t)\big){\rm d}h_{\ell}(t)>0$. 
Then, by Proposition \ref{prop-Cthm}, we know that the sufficient and necessary condition for the dominance of $F_T$ over $G_T$ within the class $\mathcal{U}_{1+\ell}$ is  
\begin{equation*}\label{sncondition}
	\int_{0}^{Z_{\Lambda_{p_0}}}\big(G_T(t)-F_T(t)\big){\rm d}h_{\ell}(t)\ge \int_{Z_{\Lambda_{p_0}}}^{+\infty}\big(F_T(t)-G_T(t)\big){\rm d}h_{\ell}(t),
\end{equation*}
which can be rewritten as  $e^{\frac{T}{\ell}\big(\mu_F+\frac{\sigma^2_F}{2\ell}\big) }- e^{\frac{T}{\ell}\big(\mu_G+\frac{\sigma^2_G}{2\ell}\big) }\ge 0$. This implies $F_T\succsim_{1+\ell-\textit{SD}}G_T$.

($b$). {\bf Necessity}. 
To establish the necessity, we aim to derive contradictions with the dominance relation $F_T\succsim_{1+\ell-\textit{SD}}G_T$ if any of the conditions in Theorem \ref{fractional} is violated. Firstly, assume that $\mu_F<\mu_G$. In this case, $G_T$
is preferred to $F_T$ under the utility function $U(x)=\log x$. Since $\log x\in\mathcal{U}_{1+\ell}$ for all $0<\ell\le 1$, it follows that $F_T$ does not stochastically dominate $G_T$ within the class  $\mathcal{U}_{1+\ell}$. 

Next, suppose that the second condition is violated, i.e., $\sigma_F>\sigma_G$. We will show that neither $F_T$ nor $G_T$ stochastically  dominates the other within the class $\mathcal{U}_{1+\ell}$. Given $\sigma_F> \sigma_G$ and $\mu_F\ge\mu_G$,   there exists a non-positive value  $Z_{p_0}=\sqrt{T}(\mu_F-\mu_G)/(\sigma_G-\sigma_F)$ such that the equality \eqref{equ} holds.  
For any value $Z_{p}<Z_{p_0}$, we have that $Z_{\Lambda_p}(F_T)<Z_{\Lambda_p}(G_T)$ and $F_T$ is above  $G_T$. Conversely, for any value $Z_{p}>Z_{p_0}$, we have that  $Z_{\Lambda_p}(F_T)>Z_{\Lambda_p}(G_T)$ and $F_T$ is below  $G_T$. Consequently, $F_T$ and $G_T$ are single crossing at $Z_{\Lambda_{p_0}}$. Then, for all $t\le Z_{\Lambda_{p_0}}$,  $$\int^{t}_{0}\big(G_T(t)-F_T(t)\big){\rm d}h_{\ell}(t)<0,$$ 
so the inequality \eqref{cfracSD} fails to hold. This implies that $F_T$ does not stochastically  dominate $G_T$ within $\mathcal{U}_{1+\ell}$.  
Also, since 
$\sigma_F> \sigma_G$ and $\mu_F\ge\mu_G$ by assumptions, then $\mathbb{E}_{G}[W_T]=e^{T(\mu_G+\sigma^2_G/2)}<e^{T(\mu_F+\sigma^2_F/2)}=\mathbb{E}_{F}[W_T]$, and thus $G_T$ does not stochastically  dominate $F_T$ within $\mathcal{U}_{1+\ell}$. 

Finally, we verify the inequality $\mu_F+\frac{\sigma^2_F}{2\ell}\ge \mu_G+\frac{\sigma^2_G}{2\ell}$. If  $\mu_F+\frac{\sigma^2_F}{2\ell}< \mu_G+\frac{\sigma^2_G}{2\ell}$, then we choose the utility function $U(x)=\ell x^{1/\ell}$, which belongs to the class $\mathcal{U}_{1+\ell}$. In this case, \begin{equation}\label{ell3}
	\mathbb{E}_F[U(W_T)]-\mathbb{E}_G[U(W_T)]=\ell\big(\mathbb{E}_F[W_T^{1/\ell}]-\mathbb{E}_G[W_T^{1/\ell}]\big)=\ell\big(e^{(\frac{\mu_F+\sigma^2_F/(2\ell)}{\ell})T}-e^{(\frac{\mu_G+\sigma^2_G/(2\ell)}{\ell})T}\big)<0,
\end{equation} which is a contradiction with  the dominance relation  $F_T\succsim_{1+\ell-\textit{SD}}G_T$.

\subsection{Proof of Theorem \ref{thm:AFSD}}

	($a$). {\bf Necessity}. We aim to derive a contradiction with the asymptotic dominance relation $F_T\succsim_{1+\ell-\textit{ASD}}G_T$  introduced in Definition \ref{AFSD} if the condition \eqref{condition1} is violated. If, by contradiction $\sigma_F>\sigma_G$, one can choose a constant  $\alpha<2\min\{\frac{\mu_G-\mu_F}{\sigma^2_F-\sigma^2_G},0,-\frac{\mu_G}{\sigma^2_G}\}$ such that $\mu_F+\frac{\alpha}{2}\sigma^2_F<\mu_G+\frac{\alpha}{2}\sigma^2_G<0$. 
	
	Let $U(t)=\frac{t^{\alpha}}{\alpha}$, $t\ge0$. Since $\alpha<0$, then for any $0<\ell< 1$, $U'(t)/t^{1/\ell-1}=t^{\alpha-1/\ell}$ is non-increasing on $(0,\infty)$ such that $U\in\mathcal{U}_{1+\ell}$. Noting that  $\mathbb{E}_F[\frac{W_T^{\alpha}}{\alpha}]-\mathbb{E}_G[\frac{W_T^{\alpha}}{\alpha}]=\frac{e^{\alpha(\mu_F+\frac{\alpha}{2}\sigma^2_F)T}-e^{\alpha(\mu_G+\frac{\alpha}{2}\sigma^2_G)T}}{\alpha}<0$ for any $T$, we know that  $\liminf\limits_{T\rightarrow\infty}\big(\mathbb{E}_F[U(W_T)]-\mathbb{E}_G[U(W_T)]\big)=-\infty$, which is a contradiction with  the asymptotic dominance relation  $F_T\succsim_{1+\ell-\textit{ASD}}G_T$. Hence,  it must follow that  $\sigma_F\le\sigma_G$. If the equality $\sigma_F=\sigma_G$ holds, then $\mu_F>\mu_G$ must be satisfied, because otherwise the distributions $F_T$ and $G_T$ are identical, and there does not exist a utility function such that the inequality  $\liminf\limits_{T\rightarrow\infty}\big(\mathbb{E}_F[U(W_T)]-\mathbb{E}_G[U(W_T)]\big)> 0$ holds. 
	
	If $\mu_F+\frac{\sigma^2_F}{2\ell}< \mu_G+\frac{\sigma^2_G}{2\ell}$, then we choose the utility function $U(x)=\ell x^{1/\ell}$, which belongs to the class $\mathcal{U}_{1+\ell}$. It follows from Assumption I that  $\mu_F+\frac{\sigma^2_F}{2\ell}>0$. Then, in this case,  $\liminf\limits_{T\rightarrow\infty}\big(\mathbb{E}_F[U(W_T)]-\mathbb{E}_G[U(W_T)]\big)=\ell\liminf\limits_{T\rightarrow\infty}\big(\mathbb{E}_F[W_T^{1/\ell}]-\mathbb{E}_G[W_T^{1/\ell}]\big)=\ell\liminf\limits_{T\rightarrow\infty}\big(e^{(\frac{\mu_F+\sigma^2_F/(2\ell)}{\ell})T}-e^{(\frac{\mu_G+\sigma^2_G/(2\ell)}{\ell})T}\big)=-\infty$, which is a contradiction with  the asymptotic dominance relation  $F_T\succsim_{1+\ell-\textit{ASD}}G_T$.

		($b$). {\bf Sufficiency}. Suppose that the condition \eqref{condition1} holds. Then, it follows from Theorem \ref{fractional} that \(\mathbb E_F[U(W_T)]\geq\mathbb E_G[U(W_T)]\) for every finite \(T\) and for any \(U\in\mathcal U_{1+\ell}\). Thus, \(\liminf\limits_{T\to\infty}(\mathbb E_F[U(W_T)]-\mathbb E_G[U(W_T)])\geq0\) for any  \(U\in\mathcal U_{1+\ell}\). It therefore remains to verify that there exists some  \(U\in\mathcal U_{1+\ell}\) such that \(\liminf\limits_{T\to\infty}(\mathbb E_F[U(W_T)]-\mathbb E_G[U(W_T)])>0\). In fact, 
		\[
	\mu_F-\mu_G=\left(\mu_F+\frac{\sigma_F^2}{2\ell}-\mu_G-\frac{\sigma_G^2}{2\ell}\right)+\frac{\sigma_G^2-\sigma_F^2}{2\ell}>0,
	\]
	where the inequality follows from the condition \eqref{condition1}, with at least one of the inequalities in \eqref{condition1} being strict. Let \(U(t)=\log t\), $t>0$. Notice easily that \(U\in\mathcal U_{1+\ell}\) for any $0<\ell<1$. Hence, \[\mathbb E_F[\log W_T]-\mathbb E_G[\log W_T]=T(\mu_F-\mu_G)\to+\infty\]
	as $T\to\infty$. This shows that we can choose a utility function for which the inequality \eqref{eq:AFSD} is strict. The sufficiency condition has been verified.

\subsection{Proof of Theorem \ref{thm:OAFSD}}

	($a$).	{\bf Sufficiency}. To establish the sufficiency, we first consider  the condition $\mu_F+\frac{\sigma^2_F}{2\ell}> \mu_G+\frac{\sigma^2_G}{2\ell}$.  
	
Note that any utility function \(U\in\mathcal U^*_{1+\ell}\) can be extended continuously at \(x=0\). Thus, we assume that $U(0)$ is well-defined. It follows from the condition $0<\inf\limits_{x}\frac{U'(x)}{x^{1/\ell-1}}\le \sup\limits_{x}\frac{U'(x)}{x^{1/\ell-1}}<\infty$ that for any $x\in(0,\infty)$, there exist constants \(0<\lambda_1\leq\lambda_2<\infty\) such that 
	\begin{equation}\label{E1}
		\lambda_1\ell x^{1/\ell}\leq U(x)-U(0)\leq \lambda_2\ell x^{1/\ell}.
	\end{equation}
	
	Let $Z$ be a standard normal random variable. Then
	\begin{equation}\label{eq:1}
		\mathbb{E}_F[U(W_T)]-\mathbb{E}_G[U(W_T)]=\mathbb{E}[U(e^{T\mu_F+\sqrt{T}\sigma_F Z})]-\mathbb{E}[U(e^{T\mu_G+\sqrt{T}\sigma_G Z})].
	\end{equation}
For any  \(U\in\mathcal U^*_{1+\ell}\), it follows from the inequalities in \eqref{E1} that for all $z\in \mathbb{R}$, 
$$U(e^{T\mu_F+\sqrt{T}\sigma_F z})-U(e^{T\mu_G+\sqrt{T}\sigma_G z})=U(e^{T\mu_F+\sqrt{T}\sigma_F z})-U(0)-\big(U(e^{T\mu_G+\sqrt{T}\sigma_G z})-U(0)\big) $$
	\begin{equation}\label{eq:2}\ \ \ \ \ \ \ \ \ \ \ \ \ \ \ \ \ \ \ \ \ \ \ \  \ge \lambda_1 \ell e^{\frac{(T\mu_F+\sqrt{T}\sigma_F z)}{\ell}}
	-\lambda_2 \ell e^{\frac{(T\mu_G+\sqrt{T}\sigma_G z)}{\ell}}.\end{equation}
		Combining \eqref{eq:1} and \eqref{eq:2}, it yields that 
	\begin{equation}\label{ell1}
		\mathbb{E}_F[U(W_T)]-\mathbb{E}_G[U(W_T)]\ge \int_{-\infty}^{\infty}\big(\lambda_1 \ell e^{\frac{(T\mu_F+\sqrt{T}\sigma_F z)}{\ell}}
		-\lambda_2 \ell e^{\frac{(T\mu_G+\sqrt{T}\sigma_G z)}{\ell}}\big) {\rm d}\Phi(z).  
	\end{equation}
	Let $\log(V_T)$ follow a normal distribution $N(T\mu/\ell, T\sigma^2/\ell^2)$.   Since $\mu_F+\frac{\sigma^2_F}{2\ell}> \mu_G+\frac{\sigma^2_G}{2\ell}$, then
	\begin{eqnarray}&\int_{-\infty}^{\infty}\big(\lambda_1 \ell e^{\frac{(T\mu_F+\sqrt{T}\sigma_F z)}{\ell}}
	-\lambda_2 \ell e^{\frac{(T\mu_G+\sqrt{T}\sigma_G z)}{\ell}}\big) {\rm d}\Phi(z)&=\ell(\lambda_1\mathbb{E}_F[V_T]-\lambda_2\mathbb{E}_G[V_T])\nonumber \\
&&=\ell(\lambda_1e^{\frac{T(\mu_F+\sigma^2_F/(2\ell))}{\ell}}-\lambda_2e^{\frac{T(\mu_G+\sigma^2_G /(2\ell))}{\ell}})\rightarrow+\infty\nonumber 
		\end{eqnarray}
	as $T\rightarrow\infty$. Combining this result with the inequality \eqref{ell1}, we have that \(\liminf\limits_{T\to\infty}(\mathbb E_F[U(W_T)]-\mathbb E_G[U(W_T)])=+\infty\) for any   \(U\in\mathcal U^*_{1+\ell}\).

	Next, we consider the condition  $\mu_F+\frac{\sigma^2_F}{2\ell}= \mu_G+\frac{\sigma^2_G}{2\ell}$ with $\sigma_F< \sigma_G$. Since \(\mathcal U^*_{1+\ell}\subset\mathcal U_{1+\ell}\), Theorem \ref{thm:AFSD} implies that  \(\liminf\limits_{T\to\infty}(\mathbb E_F[U(W_T)]-\mathbb E_G[U(W_T)])\geq0\) for all \(U\in\mathcal U^*_{1+\ell}\). It therefore remains to verify that there exists some utility function  \(U\in\mathcal U^*_{1+\ell}\) such that \(\liminf\limits_{T\to\infty}(\mathbb E_F[U(W_T)]-\mathbb E_G[U(W_T)])>0\).  Let  \(0<\rho<1\) be sufficiently close to 1 such that \(\mu_F+\rho\sigma_F^2/(2\ell)>0\), and define \(U_\rho(x)=\ell x^{1/\ell}+(1+x^{1/\ell})^\rho\), $x>0$. Straightforward calculations yield  $	U'_\rho(x)= x^{1/\ell-1} +\rho(1+x^{1/\ell})^{\rho-1}x^{1/\ell-1}/\ell\ge 0$ and  $U'_\rho(x)/x^{1/\ell-1}=1+\rho(1+x^{1/\ell})^{\rho-1}/\ell$, which is non-increasing on $(0,\infty)$. Moreover, 
	$1\le U'_\rho(x)/x^{1/\ell-1}\le 1+\rho/\ell$ for all $x\in(0,\infty)$. Thus, \(U_\rho\in\mathcal U^*_{1+\ell}\). 
	
	It follows from the condition \(\mu_F+\sigma_F^2/(2\ell)=\mu_G+\sigma_G^2/(2\ell)\) that
		\begin{eqnarray}\label{eq:omaintm1}
		&\mathbb{E}_F[U_\rho(W_T)]-\mathbb{E}_G[U_\rho(W_T)]&=\ell\big(\mathbb E_F[W_T^{1/\ell}]-\mathbb E_G[W_T^{1/\ell}]\big)+\mathbb E_F[(1+W_T^{1/\ell})^{\rho}] -\mathbb E_G[(1+W_T^{1/\ell})^{\rho}]\nonumber \\
		&&= \ell\big(e^{\frac{(\mu_F+\frac{\sigma^2_F}{2\ell})T}{\ell}}-e^{\frac{(\mu_G+\frac{\sigma^2_G}{2\ell})T}{\ell}}\big)+\mathbb E_F[(1+W_T^{1/\ell})^{\rho}] -\mathbb E_G[(1+W_T^{1/\ell})^{\rho}] \nonumber \\
		&&= \mathbb E_F[(1+W_T^{1/\ell})^{\rho}] -\mathbb E_G[(1+W_T^{1/\ell})^{\rho}].
	\end{eqnarray}
By the inequality \begin{equation}\label{ineq}
y^\rho\leq(1+y)^\rho\leq1+y^\rho,\ \ y\geq0,
\end{equation}
the equation \eqref{eq:omaintm1} can be further rewritten as
\begin{equation}\label{eq:omaintm2}
	\mathbb{E}_F[U_\rho(W_T)]-\mathbb{E}_G[U_\rho(W_T)]\ge \mathbb E_F[W_T^{{\rho/\ell}}]-\mathbb E_G[W_T^{{\rho/\ell}}]-1=\big(e^{\frac{\rho(\mu_F+\frac{\rho\sigma^2_F}{2\ell})T}{\ell}}-e^{\frac{\rho(\mu_G+\frac{\rho\sigma^2_G}{2\ell})T}{\ell}}\big)-1.
\end{equation}	
Also, from the condition $\sigma_F< \sigma_G$, it follows 
that 
	\[
	\frac{\rho}{\ell}\left(\mu_F+\frac{\rho\sigma_F^2}{2\ell}\right)
	-\frac{\rho}{\ell}\left(\mu_G+\frac{\rho\sigma_G^2}{2\ell}\right)
	=\frac{\rho(1-\rho)}{2\ell^2}(\sigma_G^2-\sigma_F^2)>0 .
	\]
Combining this result with the inequality \eqref{eq:omaintm2}, it yields that $\mathbb{E}_F[U_\rho(W_T)]-\mathbb{E}_G[U_\rho(W_T)]\rightarrow+\infty$ as $T\rightarrow\infty$. This shows that it is possible to determine a utility function $U$ for which \(\liminf\limits_{T\to\infty}(\mathbb E_F[U(W_T)]-\mathbb E_G[U(W_T)])>0\). 

In both cases, the sufficiency condition has been verified.

	($b$). {\bf Necessity}. To establish the necessity, we provide a proof by contradiction. 
	
	Indeed, if $\mu_F+\frac{\sigma^2_F}{2\ell}< \mu_G+\frac{\sigma^2_G}{2\ell}$, then DMs with the utility function $U(t)=\ell t^{1/\ell}\in \mathcal{U}^*_{1+\ell}$ would strictly prefer $G_T$ to $F_T$ as $T\rightarrow\infty$ according to Equality \eqref{ell3}. This contradicts the asymptotic dominance relation that $F_T$ dominates $G_T$ by general $1+\ell$-ASD. 
	
	If $\mu_F+\frac{\sigma^2_F}{2\ell}= \mu_G+\frac{\sigma^2_G}{2\ell}$, we distinguish between two distinct cases.
	
	\textit{Case 1.} $\sigma_F=\sigma_G$. In this case, $\mu_F=\mu_G$, and the distributions $F_T$ and $G_T$ are therefore identical. Consequently, there exists no utility function $U$ for which \(\liminf\limits_{T\to\infty}(\mathbb E_F[U(W_T)]-\mathbb E_G[U(W_T)])>0\).
	
	\textit{Case 2.} $\sigma_F>\sigma_G$. In this case, we  choose the utility function \(U_\rho(x)=\ell x^{1/\ell}+(1+x^{1/\ell})^\rho\). From the arguments in the proof of sufficiency, we know that 
	  $U_\rho\in\mathcal{U}^*_{1+\ell}$. Moreover, it follows from \eqref{eq:omaintm1} and \eqref{ineq} that 
\begin{equation}\label{eq:omaintm4}
	\mathbb{E}_F[U_\rho(W_T)]-\mathbb{E}_G[U_\rho(W_T)]\le 1+\mathbb E_F[W_T^{{\rho/\ell}}]-\mathbb E_G[W_T^{{\rho/\ell}}]=1+\big(e^{\frac{\rho(\mu_F+\frac{\rho\sigma^2_F}{2\ell})T}{\ell}}-e^{\frac{\rho(\mu_G+\frac{\rho\sigma^2_G}{2\ell})T}{\ell}}\big).
\end{equation}	
In this case,
	\[
\frac{\rho}{\ell}\left(\mu_G+\frac{\rho\sigma_G^2}{2\ell}\right)
-\frac{\rho}{\ell}\left(\mu_F+\frac{\rho\sigma_F^2}{2\ell}\right)
=\frac{\rho(1-\rho)}{2\ell^2}(\sigma_F^2-\sigma_G^2)>0.
\]
This result, together with the inequality \eqref{eq:omaintm4}, implies that $\liminf\limits_{T\rightarrow\infty}\Big(\mathbb{E}_F[U_\rho(W_T)]-\mathbb{E}_G[U_\rho(W_T)]\Big)=-\infty$,  which leads to a contradiction with the asymptotic dominance relation that $F_T$ dominates $G_T$ by general $1+\ell$-ASD.

\newcommand{\DOI}[1]{Doi: \href{https://doi.org/#1}{#1}}
\bibliographystyle{elsarticle-harv}
\bibliography{fractional}

\end{document}